\documentclass[hyper]{JHEP3}

\usepackage{epsfig}
\usepackage{latexsym}
\usepackage{amsfonts}
\usepackage{amsmath}
\usepackage{amsthm}
\usepackage{amssymb}
\usepackage{amsbsy}
\usepackage{multirow}

\def\Tr				{\mathrm{Tr}}

\def\calh         {{\cal H}}

\def\calm         {{\cal M}}
\def\caln         {{\cal N}}

\newsavebox{\uuunit}
\sbox{\uuunit}
    {\setlength{\unitlength}{0.825em}
     \begin{picture}(0.6,0.7)
        \thinlines
        \put(0,0){\line(1,0){0.5}}
        \put(0.15,0){\line(0,1){0.7}}
        \put(0.35,0){\line(0,1){0.8}}
       \multiput(0.3,0.8)(-0.04,-0.02){12}{\rule{0.5pt}{0.5pt}}
     \end {picture}}

\def\be{\begin{equation}}
\def\ee{\end{equation}}
\def\bea{\begin{eqnarray}}
\def\eea{\end{eqnarray}}

\def\a{\alpha}
\def\b{\beta}
\def\g{\gamma}

\def\l{\lambda}

\def\baa{ {\bar{\alpha}} }
\def\bbb{ {\bar{\beta}} }

\def\tx{\tilde{x}}
\def\tD{\tilde{D}}
\def\tL{\tilde{\lambda}}

\title{Scaling BPS Solutions and pure-Higgs States} 
\author{Iosif Bena$^1$, Micha Berkooz$^2$, Jan de Boer$^3$, Sheer El-Showk$^1$ and Dieter Van den Bleeken$^4$
\\
\\
$^1$ Institut de Physique Th\'eorique, CEA Saclay, CNRS URA 2306,
F-91191 Gif-sur-Yvette, France\\
$^2$  Department of Particle Physics and Astrophysics, Weizmann Institute of Science, Rehovot, 76100, Israel\\
$^3 $ Institute for Theoretical Physics, University of Amsterdam
Science Park 904, Postbus 94485, 1090 GL Amsterdam, The Netherlands \\
$^4$  Bo\u{g}azi\c{c}i University Physics Departement,
34342 Bebek / Istanbul; Turkey\\
}

\abstract{Depending on the value of the coupling, BPS states of type II string theory compactified on a Calabi-Yau manifold can be described as multicenter supergravity solutions or as states on the Coulomb or the Higgs branch of a quiver gauge theory. While the Coulomb-branch states can be mapped one-to-one to supergravity states, this is not automatically so for Higgs-branch states. In this paper we explicitly compute the BPS spectrum of the Higgs branch of a three-center quiver with a closed loop, and identify the subset of states that are in one-to-one correspondence with Coulomb/supergravity multicenter states. We also show that there exist additional ``pure-Higgs'' states, that exist if and only if the charges of the centers can form a scaling solution. Using generating function techniques we compute the large charge degeneracy of the ``pure-Higgs'' sector and show that it is always exponential. We also construct the map between Higgs- and Coulomb-branch states, discuss its relation to the Higgs-Coulomb map of one of the authors and Verlinde, and argue that the pure Higgs states live in the kernel of this map. Given that these states have no obvious description on the Coulomb branch or in supergravity, we discuss whether they can correspond to a single-center black hole or can be related to more complicated horizonless configurations.}

\preprint{IPhT-T12/041
}

\begin{document}

\section{Introduction and overview}

\subsection{Motivation} 

Although the Bekenstein-Hawking formula for the entropy of a black hole is widely accepted,
only limited progress has been made in understanding, from a gravitational point of view, what
the nature of the underlying microscopic degrees of freedom is. Phrasing the question in somewhat
more general terms, we can ask what is the correct gravitational description of general states in
a given charge sector of string theory. In such a sector one typically finds both gravitational solutions
that are smooth and/or horizonless, and solutions which contain one or more black holes.
Given a state, e.g. some particular excitation of strings and branes, it is in general an open problem to identify by which of these classes of gravitational solutions it is best described. Resolving this problem amounts to understanding which microscopic degrees of freedom can be
described and distinguished by an observer having only gravitational probes at hand, and which ones cannot. 

In this paper we will report on some progress in achieving this goal, albeit in the particular
context of supersymmetric solutions of $N=2$ supergravity in four dimensions. The full enumeration
of all states in a given charge sector in this theory is unknown, but we know that it contains
the supersymmetric ground states of various quantum mechanical ``gauge theories'' of quiver type, with several different gauge groups and charged matter. 
Given an open string (gauge theoretical) description of the degrees of freedom, one often finds
that the theory possesses a Higgs branch and a Coulomb branch. When taking the decoupling limit relevant for the AdS/CFT correspondence, 
the Coulomb branch is removed from the system and only the Higgs branch remains. 
There is a refinement however: a small part of the Coulomb branch 
does survive, but it does not describe independent
degrees of freedom, rather it describes some of the states on the Higgs branch using different
variables. If the Coulomb branch does not extend to infinity (due to the presence of
a D-term), only the near-Higgs Coulomb branch remains. 
%For the quantum mechanical systems that we consider, a traditional Coulomb branch where
%constituents can be separated at arbitrarily long distances from each other often does not exist, 
%in which case the entire Coulomb branch survives the decoupling limit. 
%The entire Coulomb branch 
%does then describe states on the Higgs branch. 
To be unambiguous, whenever
we refer to ``Coulomb branch'' in the remainder of this paper, we mean the part of the near-Higgs Coulomb branch
which survives the decoupling limit. 

%\comment{I'm confused here, in section three, which holds before taking any kind of decoupling limit, we recover the whole Coulomb branch of the theory in the Higgs branch, so in that section Coulomb branch doesn't mean some kind of truncated version?? D}

It is precisely the Coulomb branch variables which are useful to obtain gravitational descriptions
of states in the theory. In fact, the Coulomb branch description of a state can be mapped directly into
a solution of the supergravity equations of motion. As we will explore in detail, the Coulomb
branch description does not in general capture the full Higgs branch. 
It would appear therefore that the remaining states in the Higgs branch are essentially inaccessible to a gravitational observer.
Whether such
states exist or not depends on the details of the quiver quantum mechanical system. It is only 
for quivers with closed loops that obey some additional conditions that the Coulomb branch
description is incomplete. As was shown in \cite{Denef2007b} for particular quiver, the number
of inaccessible states can be exponentially large, much larger than the number of states that
are accessible from the Coulomb branch. We will study the conditions under which this happens in more detail, compute
the spectrum explicitly for all three-centered quiver systems, and provide a simple
criterion in terms of the geometry of the Higgs branch to distinguish inaccessible states from accessible ones. 
%
%
%We will perform a detailed counting of inaccessible states and
%show that their number can be exponentially large, much larger than the number of states that
%are accessible. \comment{I think we need to maybe rephrase this a little, denef and moore were the first to show that in certain cases you can have exponential growth. What we do first is showing the precise conditions when there is such growth and when not, extending their result to the full scaling cone, computing the spectrum explicitely and distinguishing pure-Higgs from Coulomb in some simple geometric way. D}

In spacetime, inaccessible states appear in situations in which the solution space of a multi-centered
system possesses so-called scaling regions, where the centers approach each other arbitrarily
closely, signaling the absence of a potential barrier between the Coulomb- and Higgs-branch
descriptions before decoupling. Furthermore, in the limit where the centers are on 
top of each other (where the Coulomb
and Higgs branches meet), there appears to be an emergent conformal symmetry. 
It has been suggested in the past that one should describe the physics of this limit using
superconformal quantum mechanics \cite{Michelson:1999dx},
which might have a large number of degrees of freedom relevant
for this limit. To make this precise, one should show that a further low-energy limit
exists in the quiver quantum mechanical system and that this limit which makes it superconformal. However, 
as we will argue, this cannot be achieved because the Higgs branch theory has a mass gap and at
low energies violates superconformal invariance.

We will now present the system we consider in more detail, and give an overview of our
main results and their possible implications.

\subsection{Setup - gravity} 

As mentioned, we will restrict ourselves in this paper to the BPS sector of four-dimensional $\caln=2$ supergravity.
As this supergravity is the low-energy theory describing Calabi-Yau compactifications of type II string theory/M-theory, the BPS states can be traced back to various D/M-brane configurations which provide a microscopic picture for the black holes in this theory. A first major success was the calculation of the entropy of the D4-D0 black hole by Maldacena, Strominger and Witten \cite{Maldacena1997} from the effective CFT description of an M5-brane wrapping a large divisor in the Calabi-Yau $X$. Since this calculation essentially only relies on Cardy's formula and the central charge, it is very general, universal and stable, but at the same time also very crude. Indeed, in the Cardy regime the large majority of BPS states with D4-D0 charge can be accounted for by the single-center black hole, but there is a still rather significant minority that corresponds to the states of various multicenter configurations. On the other hand, outside of the Cardy regime this balance can change and these multicenter black holes can dominate the entropy of the single-center solution \cite{Denef2007b, Boer2008b, Bena2012}.

While this situation at first sight complicates our struggle to understand the relevant quantum states, it also presents us with interesting ways to gain insight into the relation between these states 
at small and large gravitational coupling. At large coupling these BPS `multicenter' states manifest themselves as classical supersymmetric solutions to $\caln=2$ supergravity. The most general such solutions were discovered in \cite{Denef2000, Berglund2006, Bena2006b} and they are determined by the positions  $\vec{r}_p \in
{\mathbb R}^{3}$ and charges $\Gamma_p$ of $N$ dyonic centers  subject to $N-1$ ``integrability'' or ``bubble'' equations
\begin{equation}\label{eqn:int}
	\sum_{q=1,\, q\neq p}^N \frac{\langle \Gamma_p, \Gamma_q\rangle}{r_{pq}} = \langle h, \Gamma_p \rangle
\end{equation} 
Note that there are $N-1$ equations rather than $N$ as the sum over all equations is
trivial.  The $\Gamma_p = \{p^0, p^A, q_A, q_0\}$ are charge vectors in
$H^{2n}(X)$ encoding the $\{D6, D4, D2, D0\}$ electro-magnetic charges of the centers
and $\langle \cdot , \cdot \rangle$ is the symplectic pairing of electric and
magnetic charges
\begin{equation}
	\langle \Gamma, \tilde{\Gamma} \rangle = 
	-p^0 \tilde{q}_0 +p^A \tilde{q}_A -q_A \tilde{p}^A +q_0 \tilde{p}^0 
\end{equation}
Such solutions do not generically exist for all values of the scalar moduli at infinity as they can decay at codimension-one surfaces in the moduli space known as walls of marginal stability. This phenomenon of ``wall-crossing'' has received a lot of attention recently both in $\caln=2$ gauge theories as well as supergravity (see e.g. \cite{Pioline2012} for a review and introduction to the literature).

\subsection{Setup - gauge theory}

At smaller coupling, when the gravitational backreaction of the charges can be ignored, the system is better described in terms of D-branes wrapped on the Calabi-Yau manifold, and an elegant effective description can be obtained by reducing the system to $0+1$ dimensions to obtain an $\caln=4$ quiver quantum mechanics \cite{Denef2002}. 
\FIGURE{\includegraphics[scale=1]{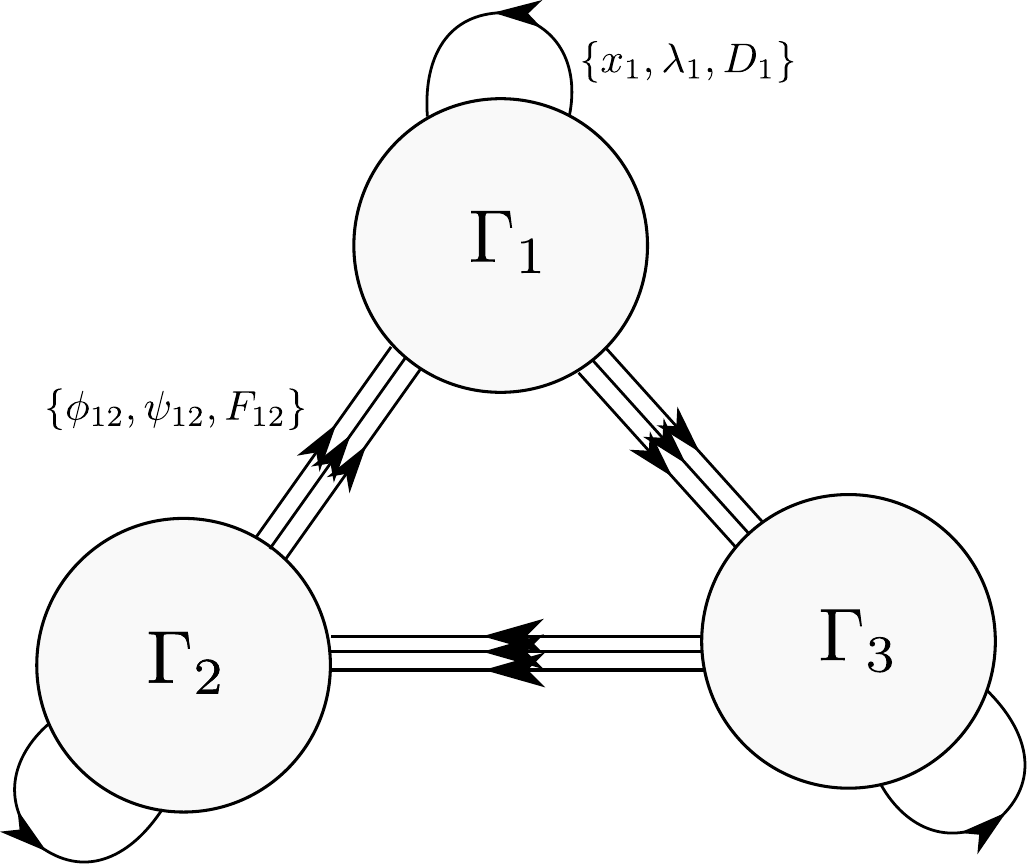}\label{pic:quiver}\caption{A three node quiver.}}

As depicted schematically in Fig. \ref{pic:quiver}, these quiver quantum theories are described in terms of two different kinds of multiplets: vector multiplets $\{\vec{x}_p, \lambda_p, D_p\}$, one for each node $\Gamma_p$ of the quiver, as well as chiral multiplets $\{\phi^\a_{pq}, \psi^\a_{pq}, F^\a_{pq} \}$ coming from the strings stretched between every pair of centers. For each pair of centers there are $\Gamma_{pq} = \langle \Gamma_p, \Gamma_q\rangle$ such chiral multiplets. The space of vacua of this quiver quantum mechanics contains both a Higgs branch and Coulomb branch, and the BPS states can be mainly supported on either branch. Moreover, this support can shift from one branch to the other as one varies the effective coupling.

One of the interesting results of \cite{Denef2002} is that for a two-center quiver one can map states directly from the Higgs branch to the Coulomb branch and to supergravity. Thus, one can basically ``follow'' a state as one turns on the gravitational coupling. A crucial observation made by \cite{Denef2002} was that, once quantum corrections are taken into account, the Coulomb branch of the quiver quantum mechanics is actually identical to the supergravity ``solution space,'' as both are parameterized by the locations of $N$ centers subject to the same constraint equations (\ref{eqn:int})! 
The profound origin of this is a non-renormalization theorem that protects the symplectic form which governs the Coulomb branch/supergravity BPS solution space and gives it a natural interpretation as a phase space. This symplectic form can in turn be used to geometrically quantize the BPS solution space \cite{Boer2009, Manschot2011}, and this corresponds physically to quantizing the angular momentum originating from the electro-magnetic interactions of the various dyonic charges \cite{Bena:2007qc}. 

On the Higgs branch the BPS states are represented as non-trivial cohomology classes on the manifold carved out by the D- and F-term constraints. One can map states on the Higgs branch to those on the Coulomb branch/supergravity by identifying the quantum numbers under the Lefschetz SU(2) to the angular momentum quantum number. This is a special example of the Higgs-Coulomb map \cite{Berkooz1999}, and for two centers this map is always one-to-one.

The situation becomes more subtle when we consider three centers. Both the Coulomb branch (which can be identified with supergravity solutions) and the Higgs branch change in an essential way. 
From the perspective of the multicenter supergravity solution the centers no longer sit at a fixed distance. Furthermore, when the intersection products satisfy the triangle inequalities ($\Gamma_{12}+\Gamma_{23}\geq \Gamma_{31}$ and cyclic), the three centers can approach each other arbitrarily close (in coordinate space) seemingly connecting the single and multicenter solution spaces. However, when one looks at the full supergravity solution one finds that in this limit the multicenter solution rather develops an infinitely deep AdS$_2$ throat. From outside the solution looks like the AdS$_2$ near-horizon region of a single-center black hole, but at the bottom of the throat the distance between the three centers remains fixed as the throat becomes longer and longer \cite{Bena2006d, Bena:2007qc, Denef2007b}.  A scaling symmetry emerges in this limit and hence these BPS solutions are often referred to as scaling solutions. The quantization of such solution spaces proceeds much as in the non-scaling case \cite{Boer2009} but one finds that quantum corrections can destroy or ``cap off'' the infinitely deep AdS$_2$ throat. This will be discussed in more detail in section~4.

On the Higgs branch, by contrast, the difference between scaling and non-scaling solutions is rather subtle.  For certain values of the charges the arrows between the nodes form a closed quiver, and in the quiver quantum mechanics this allows for the existence of a non-trivial superpotential which affects the structure of the supersymmetric states. However, not all closed quivers have intersection products that satisfy the triangle inequalities, and hence correspond to scaling supergravity solutions. 

One can count the degeneracy of the Higgs branch of the quiver quantum mechanics when the quiver is closed and has a superpotential \cite{Denef2007b}, and the result is quite remarkable: the theory has an exponential growth of states precisely when the closed quiver satisfies the triangle inequality, and there exists a scaling solution in supergravity! 
%Moreover, the growth with the charges of these states is the same as for a single-center black hole \comment{Is this the case? The standard 
%Cardy regime lies on the boundary of the scaling cone which as I understood is inaccessible to both Denef-Moore and our methods. D  }. 
These `scaling' states of the Higgs branch vastly outnumber the Coulomb-branch states, so the one-to-one map between Higgs and Coulomb branch states that we had for two centers does not hold anymore. However, the map remains surjective and one can still describe all the Coulomb-branch states on the Higgs branch, exactly as for 
non-scaling and no-superpotential quivers. 
%\comment{Rewrote last sentence and addressed Iosif's comment, ok? D}\comment{The content of this paragraph is essentially correct, but messes up the presentation a little, the original idea was to first review what denef and moore did and then in the next subsection say what we added to that, but this paragraph now already contains some of our new findings... D}

These results lead to a number of interesting questions: what happens to the map between Higgs and Coulomb branch states when one goes from weak to stronger coupling? What is the nature of the extra Higgs branch states? Can they be identified? Do they have any analogue in supergravity? Is there any relation between the scaling point and the single center black hole? How generic is the exponential growth?

%{\bf Iosif: I wrote this extra paragraph - please check if you agree and massacre at will} 
Clearly these questions are crucial for our understanding of black hole physics. If one could argue that these exponentially-growing states correspond to horizonless supergravity solutions, and we found evidence to the contrary, or to more complicated horizonless stringy configurations, this would indicate that a black-hole-like entropy can be obtained from fuzzballs, and would essentially establish that the fuzzball proposal\footnote{See \cite{Mathur:2005zp, Bena:2007kg, Mathur:2008nj, Balasubramanian:2008da, Skenderis:2008qn, Chowdhury:2010ct}
 for reviews.} applies to ${\cal N}=2$ BPS black holes. On the other hand, if the exponentially-growing states will not have any support on the Coulomb branch and cannot
be captured by more complicated closed string degrees of freedom, they will all develop a horizon and become indistinguishable from the classical black hole.

In this paper we take several steps towards answering some of these questions by re-investigating the three-center scaling quiver.

\subsection{Summary and Results}

After shortly reviewing in section \ref{sec:qqm} the general structure of quiver quantum mechanics, we specialize to a three-node quiver with a closed loop and generic superpotential in section \ref{sec:cohomology}. Following \cite{Denef2007b} we review how the Higgs branch is a complete intersection manifold. We then apply the Lefschetz hyperplane theorem to compute its Betti-numbers, and hence the BPS spectrum. It turns out that the Higgs cohomology consists of states with non-vanishing Lefschetz SU(2) quantum numbers, which map bijectively to states on the Coulomb branch. However, in the middle cohomology there are additional classes, all in the trivial representation of SU(2), that have no counterpart on the Coulomb branch. We will refer to these states as {\em pure-Higgs states}.
If one would want to add them to the Coulomb branch ``by hand'', their quantum numbers correspond to zero angular momentum, suggesting that they should indeed be related to the scaling point or to the single center black hole. 

%We will argue however that the scaling point itself is unreliable, and therefore the  pure-Higgs states have to belong to the single center black hole. {Iosif: this is not the case : the scaling point is not reliable BECAUSE we only look at Coulomb DOF. If we had the higgs DOF as well, the quantum phase space would not be so squeezed, and this scalin point could become reliable.  So the argument is circular. }

%Moreover, these states remain localized on the Higgs branch as we turn on the coupling, even when the other states on the Higgs branch move on to the Coulomb branch. {\bf Iosif:  Do we show this in the paper ? We say something else in the next subsection.}\comment{I dont think we have a rigorous proof, our argument is more like 'there is nothing left on the Coulomb branch/sugra to map them to' D} 

To proceed we compute a generating function for the  supersymmetric index  $\Omega(a,b;c)$ of three center BPS states with intersection products $a=\Gamma_{12}, b=\Gamma_{23},$
and $c=\Gamma_{31}$:
\be
Z_\Omega=\frac{x y(1-x y)}{(1+x)^2 (1+y)^2 (1-x y-x z-y z-2 x y z)} = \sum_{a,b,c=0}^\infty \Omega(a,b; c) x^a y^b z^c
\ee

One of the key results of our investigation is that this function is not symmetric in the
pairings $a,b$ and $c$ precisely because of the Higgs-branch states that map to the Coulomb
branch!  Indeed, we can compute the spectrum explicitly, and isolate the pure-Higgs states from those that have a Coulomb interpretation and count their number $\beta(a,b,c)$ separately. Their generating function is
\be
Z_\beta=\frac{x^2 y^2 z^2}{(1-x y) (1-x z) (1-y z) (1-xy-yz-zx-2 x y z)}\,.
\ee

From this generating function we can learn a number of things.
\begin{itemize}
\item $\beta$ is non-vanishing iff $a+b-2\geq c$ and cyclic, and hence pure-Higgs states only exist when the Coulomb branch has a scaling point, and viceversa.
\item for any $a,b,c\gg 1$ that satisfy the triangle equations the number of states has an exponential growth, that we calculate precisely 
\be
\beta(a,b,c)\sim\frac{2}{\pi}\sqrt{\frac{abc(ABC)^3}{(aA+bB+cC)^7}}\frac{a^ab^bc^c}{A^AB^BC^C}\,2^{a+b+c}\,,
\ee
where $A\equiv -a+b+c, B\equiv a-b+c, C\equiv a+b-c$.
\item the generating function is symmetric in $a,b,c$, suggesting that different Higgs branches can share their pure-Higgs sector, but differ in the part that maps to the Coulomb branch. Furthermore this hints that these states are everywhere stable on the moduli space.
\item the generating function is combinatoric, hinting at a simple interpretation in terms of brane/string constituents.
\end{itemize}

In section \ref{sec:hc} we discuss how the general notion of Higgs-Coulomb map of \cite{Berkooz1999} is realized in our system. This is a non-trivial extension of \cite{Berkooz1999} to  multiple interacting mutually-non-local branes, and the emergence of this map is more complicated. 
We demonstrate nonetheless how the Coulomb-branch degrees of freedom still emerge from operators on the Higgs branch, and that the Coulomb branch variables are
still fermion bilinears and hence can not be fully treated as classical bosonic variables. 

In particular, we use this intuition to explain why the scaling point is unreliable and conjecture that a similar mechanism is at work in the superconformal
quantum mechanical setup of \cite{Michelson:1999dx}. 
That the scaling point is unreliable had also been observed before from a pure gravitational point of view, when studying some puzzling aspects of
scaling BPS solutions. One such puzzle was the discrepancy between having infinitely deep smooth throats and being in a solution dual to a finite-mass-gap CFT \cite{Bena:2007qc,Bena2006d}, and the quantization of \cite{Boer2009} managed to address this by arguing that  the throats will be capped. To accomplish this required the rather remarkable claim that a large macroscopically smooth spacetime is ``destroyed'' by quantum corrections  \cite{Bena:2007qc,Boer2009}, or more precisely that all throats beyond a certain depth do not have corresponding semi-classical BPS quantum states. This result relied on quantizing only the Coulomb-branch degrees of freedom, and on the fact that the phase space of these degrees of freedom becomes very restricted in the scaling region. We revisit this issue from a different perspective in  section~4.

\subsection{Discussion and Outlook}

In this paper we have laid the groundwork for a more detailed understanding of the pure-Higgs states whose exact role in wall-crossing and black hole microstate counting remains unclear. 

Many other interesting questions present themselves:

\begin{itemize}
\item Our discussion was limited to quivers with three nodes. It would be interesting to extend the Higgs-Coulomb map to quivers with more
than three nodes with various combinations of bifundamental matter.
	\item As noted earlier the scaling point contains an AdS$_2$ factor suggesting the pure-Higgs states may indeed be the states of a putative dual CFT$_1$.  These states would seem to capture the behavior of a certain set of multi-AdS$_2$ throats inside an asymptotically AdS$_2$ region (reminiscent of \cite{Maldacena1999b}).  Clearly, it is difficult to make this precise, as the scaling point is unreliable and the AdS$_2$ decoupling limit is singular from the quiver point of view. Perhaps this is a general lesson for AdS$_2$ geometries that appear in string theory. 
Understanding this issue further may shed light on the construction of the BPS sector of a CFT$_1$ dual.
	\item As suggested above, the symmetric structure of the pure-Higgs partition function suggests that it does not decay across walls of marginal stability.  This seems in accord with the result of \cite{Sen2011} on the equivalence of Higgs and Coulomb branch wall crossing. As pure-Higgs states are not present on the Coulomb branch\footnote{Note that what we call `pure-Higgs' contributions are distinct from the `scaling contributions' discussed in \cite{Manschot2011, Manschot2011a, Sen2011}, as those are actually the contributions that do map into the Coulomb branch.} the result of  \cite{Sen2011} would also imply they cannot be involved in wall-crossing and should hence be stable on all of moduli space. This would be an additional argument to compare them to black hole microstates.    It would be interesting to further study the role of the pure-Higgs states in the ${\mathcal N}=2$ partition function.
	\item The combinatorial nature of the ``pure Higgs'' partition function strongly suggests some elegant combinatorial origin, perhaps related to fermionic degrees of freedom on strings stretched between the centers.  Finding this combinatorial explanation will likely lead to a deeper understanding of these microstates and possibly also their strict AdS$_2$ limit, if such a thing exists.
	\item It is an interesting question whether the pure-Higgs branch states
can be obtained from a Coulomb branch description of a different quiver
with the same total charges.
% perhaps in a different duality frame 
It is in principle possible to obtain substantial
numbers of states from multi-centered configurations that are $SU(2)$-invariant.
Such states are not present in three-centered quivers, but will generically
be present as one increases the number of centers\footnote{One explicit example is the ``pincer'' solution of \cite{Bena2006d}. One can also examine these states 
by following attractor flow trees of non-scaling solutions, which allow
one to keep track of their $SU(2)$-content.}. These $SU(2)$ singlets are however always accompanied by non-$SU(2)$ singlets, which have a rather similar degeneracy. Hence, if they represent the pure-Higgs states there will have to exist other families of Higgs-branch states with a similar (exponential) degeneracy but with a nontrivial angular momentum, and it is not clear whether such states exist. 
	\item An important issue, that we discuss more thoroughly in section 4.5, is whether it is somehow possible to characterize the remaining states on the Higgs branch using closed string theory. This can be done for example by computing one-point functions in these states, and comparing the results to those of a single-center black hole. If they differ, then the pure Higgs states will most likely not have a horizon in the regime of parameters where supergravity is valid, and will therefore look more like fuzzballs than black holes. 
If not, then the one-point functions of the supergravity fields will be indistinguishable from those of a black hole, but it may be that one-point functions of massive string states or higher point functions of gravitational degrees of freedom will still be capable
of probing detailed properties of the missing states. Sorting out these fascinating possibilities appears to be within calculational reach, and we plan to revisit this in the future.
	\item We would like to point out that also some more basic questions concerning scaling solutions remain. Since they fall outside of the split attractor flow conjecture \cite{Denef2000, Denef2001, Denef2007b}, one would like to find another simple, robust criteria to check for their existence. Recently this was done for two scaling non-interacting centers \cite{Gaasbeek2012}, but an analogous result for interacting scaling centers is still lacking. 
%On the microscopic side, following \cite{Denef2007b}, we were able to perform our analysis assuming a generic superpotential, but it would be interesting to find a precise calculable string theory realization of a quiver superpotential that falls into this class.
	\item Though this remains to be worked out in detail, it is tempting to conjecture that if the gravitational one-point function of the pure-Higgs states are the same as those of the single-center black hole, then most of the black hole microstates would not be accessible to gravitational observers. However, we are working in a particular duality frame and in four-dimensional supergravity, and since different duality frames and different supergravities have different gravitational observers, it is still conceivable that other gravitational observers can resolve these states.

% but it is hard to see how one could
%address this question within the framework that we have been employing.
\end{itemize}

\section{Quiver Quantum Mechanics}\label{sec:qqm}
In \cite{Denef2002} the dimensional reduction to 0+1 dimensions of the
low-energy theory living on intersecting D-branes in a Calabi-Yau $X$, was very explicitly shown to
reproduce much of the physics of multicenter BPS configurations of the ${\mathcal
N}=2$ supergravity obtained from a compactification to four dimensions on the same Calabi-Yau.  The
Lagrangian for this theory can be read from a quiver diagram (such as fig. \ref{pic:quiver})
which efficiently encodes the field content of the theory.

Every node in the quiver represents a brane of charge $\Gamma_p \in H^{2n}(X)$
and there is a corresponding vector multiplet $(\vec{x}_p, \lambda_p, D_p)$
whose bosonic component is the position of the D-brane in the three external
spatial directions.  If we allow for non-primitive charges $\Gamma_p = m
\Gamma_p'$ then each node will have an associated $U(m)$ gauge symmetry under
which the vector multiplets will be adjoint-valued.  In what follows we
restrict however to primitive vectors so $m=1$ and the vector multiplets are
uncharged.

Each pair of branes intersect $\Gamma_{pq}$ times and each intersection gives
rise to a hypermultiplet $(\phi_{pq}^\alpha, \psi_{pq}^\alpha, F_{pq}^\alpha)$
in the $U(1)\times \overline{U(1)}$ bifundamental which is represented in the
quiver as an arrow pointing from node $p$ to $q$.  The Lagrangian of the
combined system is fixed by supersymmetry and can be read off from the quiver
\cite[Appendix C]{Denef2002} 
\begin{equation}\label{eqn:lagrangian}
	\begin{split}
		{\mathcal L} = &\sum_p \frac{m_p}{2} \left( \dot{x}_p^2 + D_p^2 + 2 i \bar\lambda \dot\lambda\right) - \theta_p D_p 
		+ \sum_{q\rightarrow p} \left( |\dot\phi_{pq}|^2 + F_{pq}^2 + i \bar\psi_{pq} \dot\psi_{pq} \right) \\
		&- \sum_{q\rightarrow p} \left[ (x_{pq}^2 + D_{pq}) \phi_{pq}^2 + \bar\psi_{pq} \sigma^i x_{pq}^i \psi_{pq}  - i \sqrt{2}  ( \bar\phi_{pq} \lambda_{pq} \epsilon \psi_{pq} - h.c. ) \right]  \\
		&+ \sum_{q\rightarrow p} \left(\frac{\partial W(\phi)}{\partial \phi_{pq}^a} F^a_{pq} + h.c.\right) +
		\left(\frac{\partial^2 W(\phi)}{\partial \phi_{pq}^\a \partial \phi_{pq}^\b} \psi^\a \epsilon \psi^\b + h.c.\right)
	\end{split}
\end{equation}
where the notation $q\rightarrow p$ implies a sum over the associated arrows.
For hypermultiples we implicitly sum over ``flavor'' indices,
$\alpha=1,\cdots,\Gamma_{pq}$, so that $|\phi_{pq}|^2 =
\sum_{a=1}^{\Gamma_{pq}} \bar\phi_{pq}^\a \phi_{pq}^\a$  and for vector multiplet
components we define relative differences as $D_{pq} \equiv D_p - D_q$ and likewise for
$x_{pq}, \lambda_{pq}$.  We will not generally need many of the details of this
Lagrangian but the microscopic origin of the parameters $m_p$ and $\theta_p$
play an important role so let us recall them.  The mass $m_p$ of the D-brane
wrapping $\Gamma_p$ is given by 
\begin{equation}\label{eqn:mp}
	m_p = \frac{\sqrt{v}\, |Z(\Gamma_p)|}{g_s \ell_s}, \qquad Z(\Gamma) := \langle \Gamma, \Omega \rangle, \qquad
	\Omega := -\frac{e^{B+ iJ}}{\sqrt{\frac{4}{3} J^3}}, \qquad v = \frac{2 \, V_X}{\pi^2 \, \ell_s^6}
\end{equation}
where here $B,J \in H^2(X)$ are the IIA moduli encoding the volume of CY
cycles in units of $\ell_s$ and $V_{X}$ is the CY volume.  The Fayet-Iliopoulos
term $\theta_p$ encodes the supersymmetry preserved by the brane with respect
to the background so
\begin{equation}\label{eqn:thetap}
	\theta_p = \textrm{Im}(e^{-i\alpha} Z(\Gamma_p)) 
\end{equation}
where $e^{i\alpha} = Z(\Gamma)/|Z(\Gamma)|$ is the phase of the total charge
$\Gamma = \sum_p \Gamma_p$.  In fact it is this term which appears on the RHS of (\ref{eqn:int}) in another guise
\begin{equation}\label{eqn:htheta}
	\theta_p = \frac{1}{2} \langle h, \Gamma_p \rangle
\end{equation}
We leave a discussion of the superpotential, $W(\phi)$,
to the next section.  For more details on these quantities and our
conventions the reader may consult \cite{Denef2000, Denef2002, Boer2008b}.

The {\em classical} potential, after integrating out the auxiliary fields $D_p$
and $F_{pq}$, is of the form
\begin{equation*}
V(\vec{x}_p, \phi_{pq}) =
\sum_{p}\frac{1}{2m_{p}}\left(\theta_{p}+\sum_{p}s_{pq}|\phi_{pq}|^2\right)^2+\sum_{p<q}||(\vec{x}_p-\vec{x}_q)^2|\phi_{pq}|^2+\frac{1}{4}\left|\frac{\partial
W}{\partial \phi_{pq}}\right|^2\,,
\end{equation*}
where we have introduced the antisymmetric symbol $s_{pq} = -s_{pq} = 1$
which is positive for $q\rightarrow p$ and negative when $p\rightarrow q$.

The only {\em supersymmetric} minimum of this classical potential is the
Higgs branch, corresponding to setting $\vec{x}_{pq} = 0$ and 
\begin{equation}\label{eqn:DFterms}
	\underbrace{\sum_{p}s_{pq}|\phi_{pq}|^2 = -\theta_{p}}_{\textrm{D-term}}, \qquad
	\underbrace{\frac{\partial W}{\partial \phi_{pq}} = 0}_{\textrm{F-term}}
\end{equation}
Quantum corrections, however, modify the potential allowing for the existence
of a Coulomb branch parameterized by $\langle \vec{x}_{pq} \rangle > 0$ when
$g_s > 0$ \cite{Denef2002}.   We will not repeat the derivation of this here
but note only that the minima of this quantum corrected potential correspond to
solutions to the so-called ``integrability'' equations (\ref{eqn:int}).  Points
in the Coulomb branch are thus in one-to-one correspondence to supergravity
solutions and moreover the Coulomb branch is equivalent, as a symplectic
manifold, to the supersymmetric phase space of the corresponding family of solutions to the ${\mathcal N}=2$
supergravity theory. 

\section{Sorting out the Higgs branch}\label{sec:cohomology}
In this section we discuss the BPS states on the Higgs branch of the simplest non-trivial quiver with three nodes. This quiver was studied in  detail in \cite{Denef2007b}, where  it was shown that when the quiver has no closed loops, and hence a vanishing superpotential, the Higgs branch BPS spectrum is exactly equal to that of the Coulomb branch. More interestingly, \cite{Denef2007b} showed that when the superpotential is non-trivial, the Higgs branch can contain exponentially more states than the Coulomb branch. In the first subsection we will give a short review of their characterization of the Higgs branch as a complete intersection manifold. We then continue by computing all the individual Betti numbers, that encode the number of BPS states. Using the connection between the Lefschetz SU(2) action on cohomology and the angular momentum in space-time we can then identify those states that have an equivalent on the Coulomb branch. By subtracting those we can then isolate the 'pure-Higgs' states, i.e. those that do not appear on the Coulomb branch. As we will show these states all have zero angular momentum and their degeneracies are encoded in a very interesting and beautiful generating function. Using the combinatorics of this generating function we show that such `pure-Higgs' states are present if and only if the quantum-corrected intersection products, $\Gamma_{12}-2$, $\Gamma_{23}-2$ and $\Gamma_{31}-2$, satisfy the triangle inequalities. Under the same condition the Coulomb branch contains a scaling point and this result is thus additional evidence that the scaling point contains more micro-states than are apparent from a naive supergravity analysis. Finally, using the generating function we compute the number of 'pure-Higgs' states for large charges, generalizing the result of \cite{Denef2007b} to the whole scaling cone.

\subsection{The $a\,b\,c$ of closed-loop three-quivers}\label{subsec:abc}
To ease notation we denote the three intersection products respectively as $a=\Gamma_{12}$,
$b=\Gamma_{23}$ and $c=\Gamma_{31}$. Since we assume the quiver to have a closed loop we can label the charges in such a way that all the intersection products are positive, i.e. $a,b,c>0$. The quiver quantum mechanics can then  have a non-trivial gauge invariant
superpotential, which can consistently be assumed to contain only cubic terms \cite{Denef2007b}:
\be
W=w_{\a\b\g}\phi_{12}^\a\phi_{23}^\b\phi_{31}^\gamma
\ee
The D-term constraints
\bea
|\phi_{12}|^2-|\phi_{31}|^2=-\theta_{1},& |\phi_{23}|^2-|\phi_{12}|^2=-\theta_{2},& |\phi_{31}|^2-|\phi_{23}|^2=-\theta_{3}
\eea
and the F-term constraints coming from this superpotential : 
\bea
w_{\a\b\g}\phi_{23}^\b\phi_{31}^\gamma=0,& w_{\a\b\g}\phi_{12}^\a\phi_{31}^\gamma=0& w_{\a\b\g}\phi_{12}^\a\phi_{23}^\b=0
\eea
define the Higgs branch. As was shown in \cite{Denef2007b}, for generic coefficients $w_{\a\b\g}$, all the
solutions of these equations fall into one of three branches, characterized by
$\phi_{12}=0$, $\phi_{23}=0$ or $\phi_{31}=0$ respectively. Which branch is
selected depends on the sign of the FI-terms $\theta_{i}$. Without loss of
generality one can make the choice $\theta_1,\theta_2<0$, such that the equations imply\footnote{Remember, by definition $\theta_3=-(\theta_1+\theta_2)$.} $\phi_{31}=0$. The D-term equations then simply describe a
$\mathbb{CP}^{a-1}\times\mathbb{CP}^{b-1}$. The remaining F-term constraint
imposes an additional $c$ quadratic equations on this manifold. In mathematical terms this means that the Higgs branch is a complete intersection manifold, that we will denote by $\calm_{ab}^c$.

\subsection{Computation of the Cohomology}
As is typical for quantum mechanics with extended supersymmetry, the BPS states on the Higgs branch are given by the elements of its cohomology. Since the Higgs branch of our interest $\calm_{ab}^{c}$ is a complete intersection manifold we can actually compute its Betti numbers using the following result.

Let $Y$ be a complete intersection manifold, defined as the zero locus of $k$ polynomials in a compact complex manifold $X$ of complex dimension $d$. An iterative application of the Lefschetz hyperplane theorem implies that there exists a positive integer $\beta\geq0$ such that
\be
b^i(Y)=\begin{cases} b^i(X)\quad&\mbox{when}\quad i<d-k\\
b^{d-k}(X)+\beta&\mbox{when}\quad i=d-k\\
b^{2k+i}(X)\quad&\mbox{when}\quad d-k<i
\end{cases}\label{LHT}
\ee
In words, this mathematical result states that every cohomology class of $Y$ originates from a cohomology class of the ambient manifold $X$, except for the middle cohomology of $Y$, where there can be additional classes, not related to the cohomology of $X$.

For our Higgs branch we have that $\calm_{ab}^c=Y$ for $X=\mathbb{CP}^{a-1}\times\mathbb{CP}^{b-1}$ and $k=c$, so it follows that
\be
b^i(\calm_{ab}^c)=\begin{cases}b^i\left(\mathbb{CP}^{a-1}\times\mathbb{CP}^{b-1}\right)&\mbox{when}\quad i<a+b-c-2\\
b^{a+b-2-c}\left(\mathbb{CP}^{a-1}\times\mathbb{CP}^{b-1}\right)+\b(a,b;c)&\mbox{when}\quad i=a+b-c-2\\
b^{2c+i}\left(\mathbb{CP}^{a-1}\times\mathbb{CP}^{b-1}\right)&\mbox{when}\quad a+b-c-2<i
\end{cases}\label{coh:higgs}
\ee

The cohomology of $\mathbb{CP}^{a-1}\times\mathbb{CP}^{b-1}$ is rather simple as we will review now. While doing so we will also point out how it decomposes under the Lefschetz representation of SU(2), as this will be useful in the next subsection where we discuss the relation to states on the Coulomb branch. The Lefschetz representation on the cohomology $H^\star$ of a K\"ahler manifold of complex dimension $d$, is defined by identifying $J_+=\omega\wedge\,,\ J_-=i(\omega,\cdot)\,,\ J_z=\frac{\deg-d}{2}$ and hence $|J|=\frac{n}{2}$, where $\omega$ is the K\"ahler form and $\deg$ is the degree of a class.

\FIGURE{\includegraphics[scale=0.3]{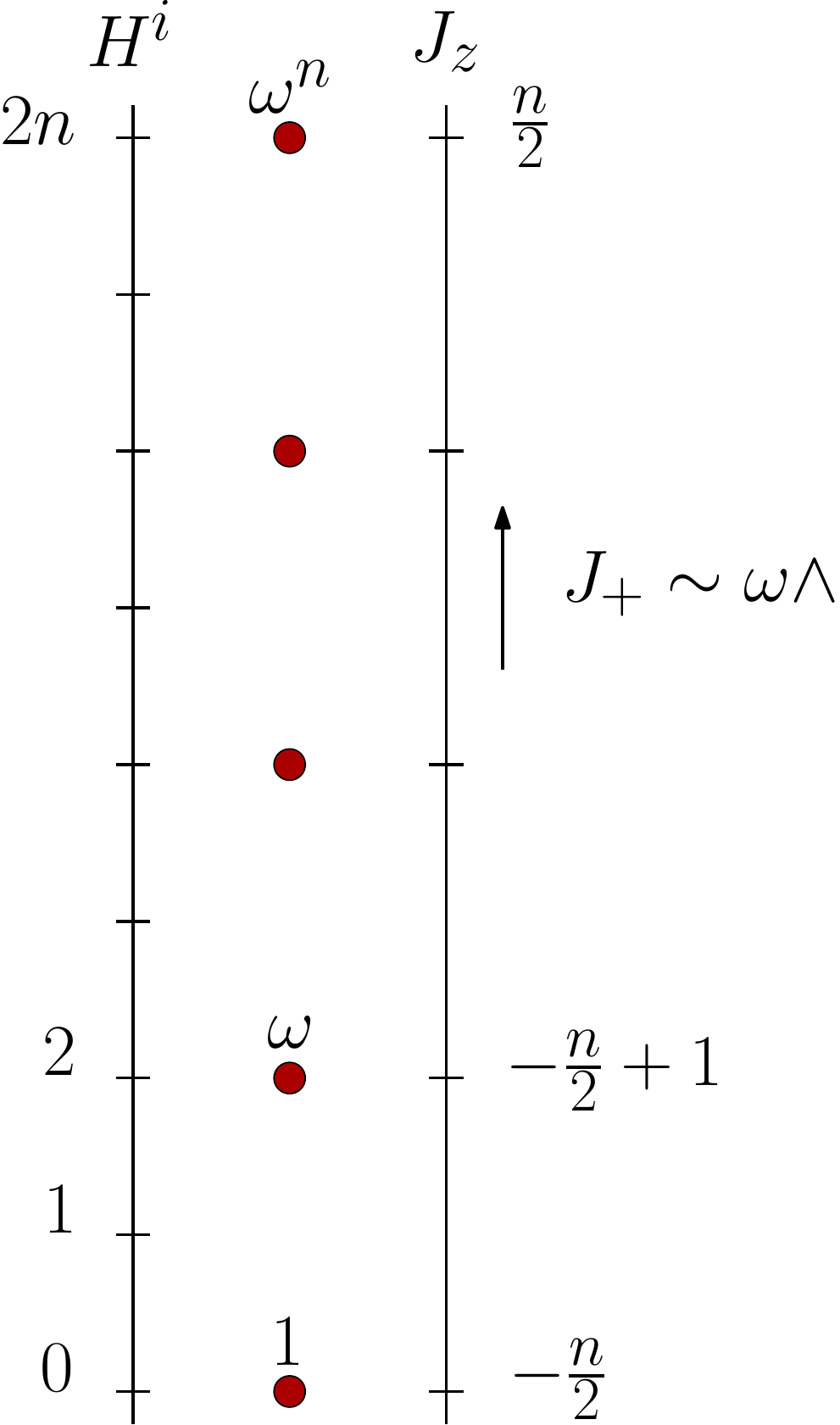}\label{pic:cpn}\caption{The cohomology of $\mathbb{CP}^n$ as the spin $\frac{n}{2}$ Lefschetz representation of SU(2).}}

The first step is to recall the cohomology structure of $\mathbb{CP}^{n}$, whose only non-trivial cohomology classes are the K\"ahler form and its products: $\omega, \omega^2\ldots, \omega^n$. Hence, the odd cohomology groups are all zero-dimensional and the even cohomology groups have dimension one: 
\be
b^{2i+1}\left(\mathbb{CP}^n\right)=0\, \qquad\qquad\qquad b^{2i}\left(\mathbb{CP}^n\right)=1 \quad(\mbox{when}\quad i\leq n)\nonumber
\ee  This is also the simplest example to illustrate the Lefschetz representation of SU(2), as $H^\star(\mathbb{CP}^n)$ corresponds to a single spin $\frac{n}{2}$ representation.  Indeed, the only states present are the $n+1$ states created by acting with the raising operator $J_+=\omega\wedge$ on the unique lowest angular momentum state $1\in H^0$. We have illustrated this in figure \ref{pic:cpn}.

The cohomology of $\mathbb{CP}^{n}\times\mathbb{CP}^{m}$ is simply given by the tensor product. As familiar, the tensor product of a spin $\frac{n}{2}$ and a spin $\frac{m}{2}$ representation  decomposes into a sum of irreducible spin $\frac{|n-m|}{2}$ to $\frac{n+m}{2}$ representations. From the point of view of cohomology this is realized as follows:  we can generate new non-trivial classes by wedging with either $\omega_1$ or $\omega_2$, but the raising operator of SU(2) is actually $\omega\wedge=(\omega_1+\omega_2)\wedge$, so we will have different highest-weight states. A simple counting per degree (see for example figure \ref{pic:cpcp}) gives the following Betti numbers (when $n\leq m$):
\bea
b^{2i+1}\left(\mathbb{CP}^n\times\mathbb{CP}^m\right)&=&0\nonumber\\
b^{2i}\left(\mathbb{CP}^n\times\mathbb{CP}^m\right)&=&
\begin{cases}i+1&\mbox{when}\quad i\leq n-1\\
n+1&\mbox{when}\quad n\leq i \leq m\\
m+n+1-i&\mbox{when}\quad m+1\leq i\leq n+m
\end{cases}\label{coh:cpcp}
\eea

\FIGURE{\includegraphics[scale=0.8]{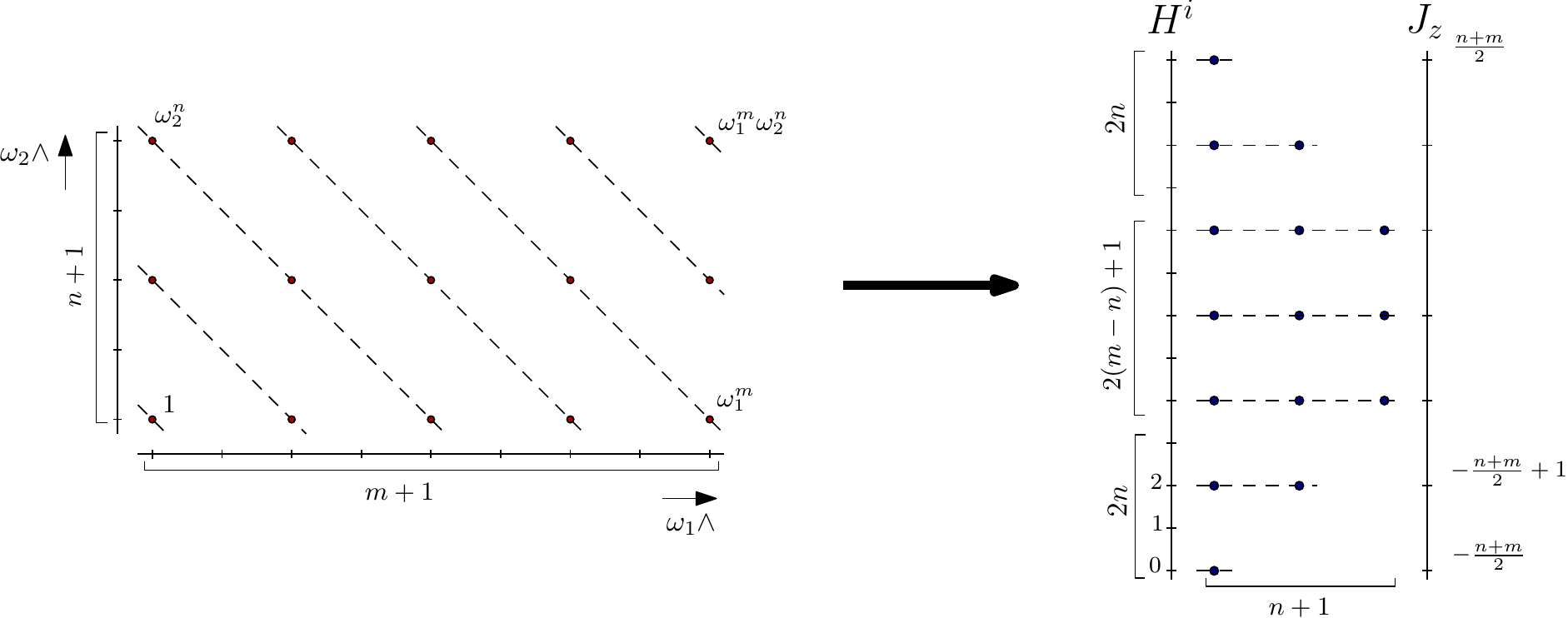}\label{pic:cpcp}
\caption{The cohomology of $\mathbb{CP}^n\times \mathbb{CP}^m$ as the spin $\frac{n}{2}\otimes\frac{m}{2}=\oplus_{j=|m-n|}^{m+n}\frac{j}{2}$\ Lefschetz representation of SU(2). Note that as cohomology classes the (blue) dots on the right, organized vertically into irreducible SU(2) representations, are linear combinations of the (red) dots on the left that are simply powers of $\omega_1$ and $\omega_2$. }}

Combining (\ref{coh:higgs}) and (\ref{coh:cpcp}) gives us, up to the precise value of $\beta$, complete knowledge of the cohomology groups of the Higgs branch $\calm_{ab}^c$. In the next two subsections we will further analyze this result: First, we will argue that those cohomology classes originating form classes on $\mathbb{CP}^{a-1}\times\mathbb{CP}^{b-1}$ are in one to one correspondence to states on the Coulomb branch. Second, we will show when there are additional states, by computing $\beta(a,b,c)$ and determining the conditions under which it is strictly positive. Third, we will give a precise estimate for this number when $a,b$ and $c$ are large.

\subsection{Emergence of the Coulomb branch}
As shown in the previous subsection, the states on the Higgs branch $\calm_{ab}^c$ fall into two classes. First there are those that originate from the cohomology of $\mathbb{CP}^{a-1}\times\mathbb{CP}^{b-1}$, and secondly there are an additional $\beta(a,b,c)$ states in the middle cohomology of $\calm_{ab}^c$. It turns out that the first class can be identified with the states on the Coulomb branch, while the $\beta(a,b,c)$ additional states have no such interpretation and will therefore be referred to as `pure-Higgs' states.

To make this identification we will show that the cohomology classes originating from $\mathbb{CP}^{a-1}\times\mathbb{CP}^{b-1}$ do not mix with the pure-Higgs states under the Lefschetz SU(2) and form an independent spin $j_1\otimes j_2$ representation, exactly as the states on the Coulomb branch. For the Coulomb branch this was shown in \cite{Boer2009}, where $j_1=\frac{j_++j_--1}{2}$ and $j_2=\frac{j_+-j_--1}{2}$, with $j_+/j_-$ the maximal/minimal {\it classical} size of the angular momentum realized on the Coulomb branch. Note that the total number of such states, that are either all fermionic or bosonic, is then $N=j_+^2-j_-^2$. To obtain this result from the Higgs branch, it is easiest to consider four different combinatorial situations: bosonic/fermionic and scaling/non-scaling.  

Since our whole discussion so far has been symmetric in $a$ and $b$, we can choose $a\geq b$ without loss of generality. A first difference between scaling and non-scaling quivers shows up in the Higgs branch, as it inherits its cohomology from $\mathbb{CP}^{a-1}\times\mathbb{CP}^{b-1}$ up to a degree $k=a+b-2-c$. Since the growth of cohomology by degree undergoes a change at $i=2a-2$ for $\mathbb{CP}^{a-1}\times\mathbb{CP}^{b-1}$, see (\ref{coh:cpcp}), we need to distinguish between $k\leq 2a-2$ and $k>2a-2$. This translates as $b\leq a+c$ and $b>a+c$, exactly the difference between the scaling and non-scaling regime. Furthermore, since the cohomology $\mathbb{CP}^{a-1}\times\mathbb{CP}^{b-1}$ is non-vanishing only for even degree, we need to distinguish between $k$ even or odd.

The different situations are depicted in figure \ref{htoc}, and a count of the total number of states gives:
\begin{itemize}
\item {\bf $\mathbf{b\leq a+c}$, $\mathbf{a+b-c}$ even}
\bea
N(a,b,c)&=&2\sum_{i=0}^{\frac{a+b-c}{2}-2}(i+1)+\frac{a+b-c}{2}=\frac{(a+b-c)^2}{4}\\
&=&j_+^2-j_-^2,\qquad\mbox{with }j_+=\frac{a+b-c}{2}\,,\ j_-=0
\eea

\item {\bf $\mathbf{b\leq a+c}$, $\mathbf{a+b-c}$ odd}
\bea
N(a,b,c)&=&2\sum_{i=0}^{\frac{a+b-c-1}{2}-1}(i+1)=\frac{(a+b-c-1)(a+b-c+1)}{4}\\
&=&j_+^2-j_-^2,\qquad\mbox{with }j_+=\frac{a+b-c}{2}\,,\ j_-=\frac{1}{2}
\eea

\item {\bf $\mathbf{b> a+c}$, $\mathbf{a+b-c}$ even}
\bea
N(a,b,c)&=&2\sum_{i=0}^{a-1}(i+1)+(b-a-c-2)a+a=a(b-c)\\
&=&j_+^2-j_-^2,\qquad\mbox{with }j_+=\frac{a+b-c}{2}\,,\ j_-=\frac{b-a-c}{2}
\eea

\item {\bf $\mathbf{b>a+c}$, $\mathbf{a+b-c}$ odd}

\bea
N(a,b,c)&=&2\sum_{i=0}^{a-1}(i+1)+(b-a-c-1)a=a(b-c)\\
&=&j_+^2-j_-^2,\qquad\mbox{with }j_+=\frac{a+b-c}{2}\,,\ j_-=\frac{b-a-c}{2}
\eea
\end{itemize}

\FIGURE{\begin{tabular}{|l|r|}
\hline
\includegraphics[scale=0.6]{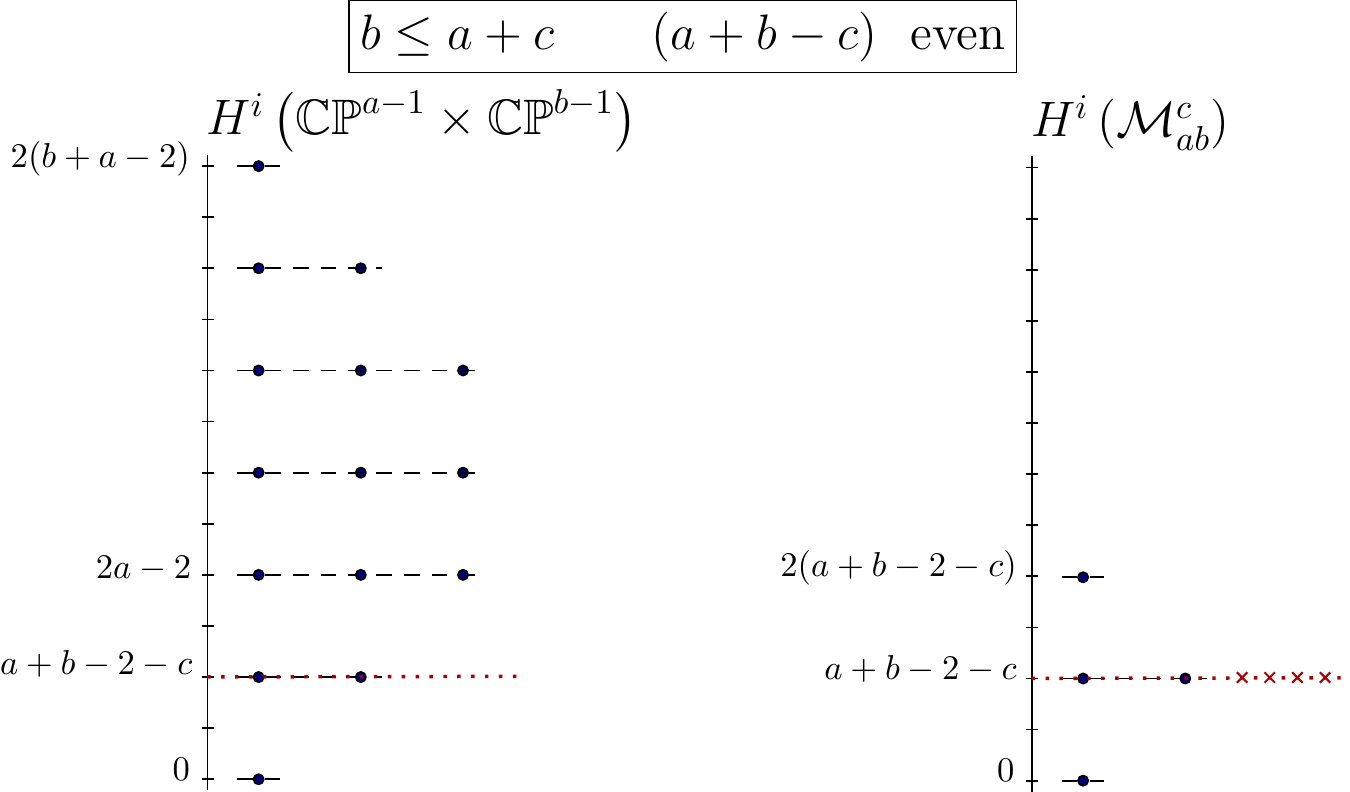}&
\includegraphics[scale=0.6]{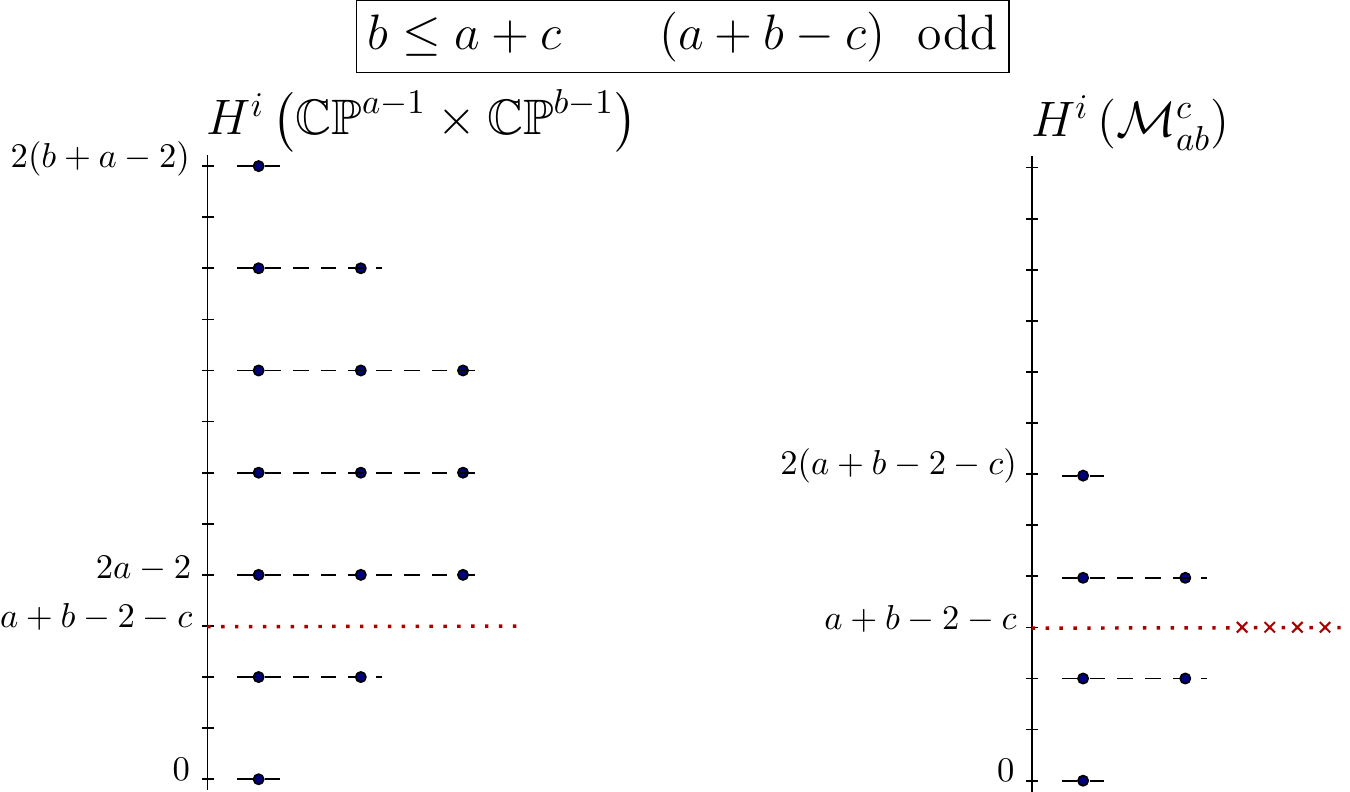}\\
\hline
\includegraphics[scale=0.6]{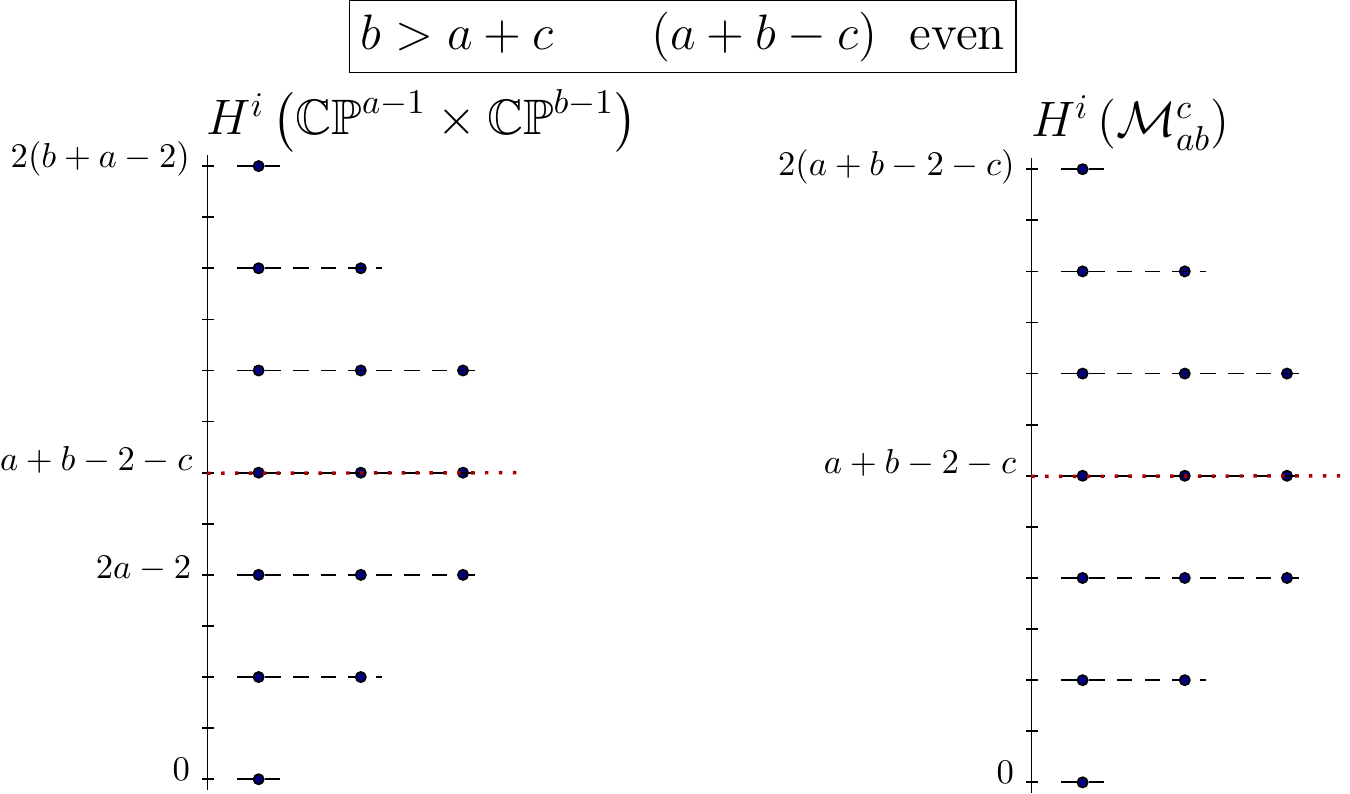}&
\includegraphics[scale=0.6]{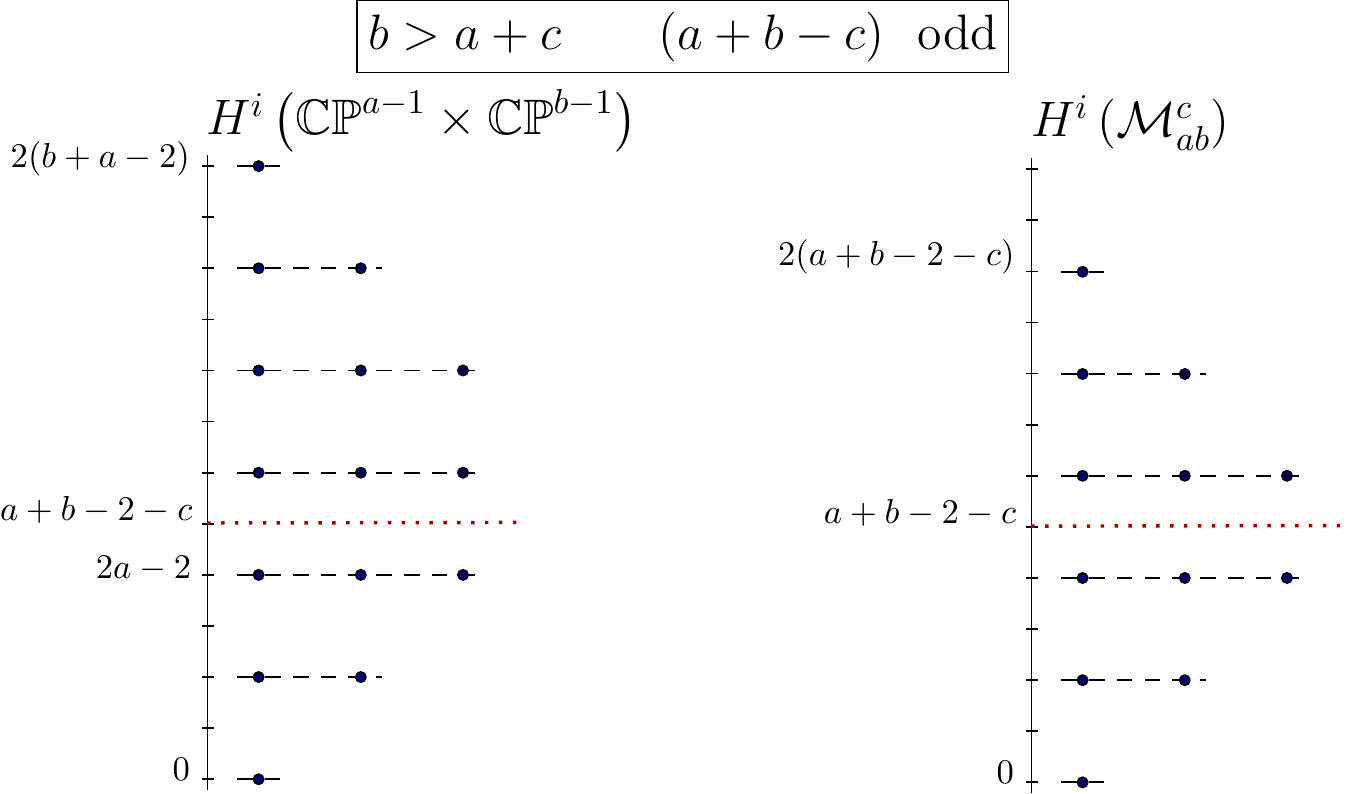}\\
\hline
\end{tabular}\caption{This figure depicts the relation between the cohomology of $\mathbb{CP}^{a-1}\times\mathbb{CP}^{b-1}$ and $\calm_{ab}^c$, given by (\ref{coh:higgs}) and (\ref{coh:cpcp}). The combinatorics gives four different situations. The (red) crosses denote the possible 'pure-Higgs' states in the middle cohomology, that are discussed in section \ref{subsec:pure}.}\label{htoc}}

The cohomologies in these four situations match perfectly with the results obtained from the Coulomb branch \cite{Boer2009}. Since they also match as SU(2) representations, we can map each  state on the Higgs branch to its unique corresponding state on the Coulomb branch, that has the same SU(2) quantum numbers. From now on we will refer to these states as 'Coulomb states', and in summary, their degeneracy is
\be
N(a,b;c)=
\begin{cases}
0&\mbox{when}\quad c>a+b-2\\
\frac{(a+b-c)^2}{4}&\mbox{when}\quad |a-b|\leq c\leq a+b-2,\ \ a+b+c \mbox{ even}\\
\frac{(a+b-c)^2-1}{4}&\mbox{when}\quad |a-b|\leq c\leq a+b-2,\ \ a+b+c \mbox{ odd}\\
a(b-c)&\mbox{when}\quad c<b-a,\ \ a<b\\
b(a-c)&\mbox{when}\quad c<a-b,\ \ b<a
\end{cases}\label{NC}
\ee

\subsection{Distillation of the pure-Higgs states}\label{subsec:pure}
We now turn our gaze to the pure-Higgs states: the states on the Higgs branch that are not in one-to-one correspondence to states on the Coulomb branch. We know they are all elements of the middle cohomology, and so carry zero angular momentum from a space-time perspective. But up to now we have only shown that their degeneracy $\beta(a,b,c)$ can be positive. Since apart from $\beta$ all the Betti numbers are known, we can compute it from the Euler characteristic $\chi$. 

Let us make this precise, and at the same time point out the relation of the different quantities to the supersymmetric index. The index of interest is the second helicity supertrace, which after factoring out the center of mass half-hypermultiplet is given by
\be
\Omega=\Tr (-1)^{2J_z}.
\ee
Under the identification of Lefschetz SU(2) with angular momentum, the z-component of the angular momentum is equal to half the difference between the degree of the form and the complex dimension of the manifold. This implies that
\be
\Omega=(-1)^{a+b+c}\chi
\ee
As we we saw in the previous subsections, up to some contribution $\beta$ at $L_z=0$, all of the cohomology classes have even degree, so we have
\bea
\Omega(a,b;c)&=&(-1)^{a+b+c}N(a,b;c)+\beta(a,b,c)\\
\beta(a,b,c)&=&(-1)^{a+b+c}(\chi(a,b;c)-N(a,b;c))\label{formb}
\eea

We computed $N(a,b;c)$ in the last subsection and because the Higgs branch $\calm_{ab}^c$ is a complete intersection manifold, there exist algebraic tools to compute its Euler characteristic. This was first done in \cite{Denef2007b}:
\be
\chi(a,b;c)=\chi\left(\calm_{ab}^c\right)=\oint dJ_1\,\oint dJ_2 \left(\frac{J_1}{1+J_1}\right)^{-a}\left(\frac{J_2}{1+J_2}\right)^{-b}\left(\frac{J_1+J_2}{1+J_1+J_2}\right)^{-c}\label{EC}
\ee

So one, albeit rather obscure, form of $\beta(a,b,c)$ is given by simply combining (\ref{NC}, \ref{EC}) and (\ref{formb}). However, a much nicer form that reveals some interesting properties of $\beta$ can be obtained by computing it's generating function, as we will now show. 

For a function $f(a,b;c)$, the generating function is defined as
\be
Z_f=\sum_{a,b,c=0}^{\infty}f(a,b;c)x^ay^bz^c\,,
\ee
and Eq. (\ref{formb}) implies that
\be
Z_\beta(x,y,z)=Z_{\chi}(-x,-y,-z)-Z_{N}(-x,-y,-z)\label{difZ}
\ee

Summing the Coulomb degeneracies (\ref{NC}) gives
\be
Z_N=\frac{x y (1-x y+ (2-x-y)x y z)}{(1-x)^2 (1-y)^2 (1-x y) (1-x z) (1-y z)}\label{ZC}
\ee 
To compute the generating function of the Euler characteristic (\ref{EC}) it is convenient to perform a change of variables in the integral. By defining $x=\frac{J_1}{1+J_1}$ and $y=\frac{J_1}{1+J_1}$ it follows that
\be
\chi(a,b;c)=
\oint dx \oint dy \,\frac{x^{-a} y^{-b}\, \left(y(1-x)+x(1-y)\right)^c}{(1-x)^2(1-y)^2(1-xy)^c}
\ee
By Cauchy's theorem these are simply the coefficients of the following meromorphic functions
\be
f_c(x,y)=\sum_{a,b=0}^\infty\chi(a,b;c)x^ay^b=\frac{x y\, \left(y(1-x)+x(1-y)\right)^c}{(1-x)^2(1-y)^2(1-xy)^c}
\ee
We can now perform the sum over $c$ by hand using the formula $\sum q^c z^c=\frac{1}{1-q z}$, the result is
\be
Z_\chi=\frac{x y(1-x y)}{(1-x)^2 (1-y)^2 (1-x y-x z-y z+2 x y z)}\label{ZE}
\ee
Combining (\ref{difZ}), (\ref{ZC}) and (\ref{ZE}) one finds the generating function of pure-Higgs states:
\be
Z_\beta=\frac{x^2 y^2 z^2}{(1-x y) (1-x z) (1-y z) (1-xy-yz-zx-2 x y z)}\label{ZB}
\ee
This equation is the central result of this paper, and in the subsections we discuss some of the physics it implies.

\subsection{Combinatorics of the pure-Higgs states}
We obtained a closed form expression for $\beta(a,b,c)$ through its generating function (\ref{ZB}) in the last subsection. Although we can now find its value for arbitrary values of $(a,b,c)$, by simply expanding the series up to sufficient order, this does not give much insight on the origin and physical interpretation of these states. In this subsection we will discuss a combinatoric interpretation of $\beta$, that will allow us to show that pure-Higgs states are present if and only if $(a-2), (b-2)$ and $(c-2)$ satisfy the triangle inequalities.

The first interesting property of $\beta(a,b,c)$, that can be inferred directly from its generating function, is that it is symmetric in $a,b$ and $c$. This is a very non-trivial fact, since the Higgs branch $\calm_{ab}^c$, its Euler characteristic\footnote{Note that in \cite{Denef2007b} it ws observed that the Euler characteristic is almost symmetric, i.e. it can be written as $\chi(a,b;c)=ab-f(a,b,c)$ with $f$ symmetric in $a,b$ and $c$. The function $f$ is however only indirectly related to $\beta$. On can define $g(a,b,c)=ab-N(a,b;c)$, which is completely symmetric as well, and thus $\beta=(-1)^{a+b+c}\left(g-f\right)$} $\chi(a,b;c)$ and the Coulomb degeneracies $N(a,b;c)$ are only symmetric in $a$ and $b$. This anti-symmetry between $a,b$ and $c$ followed from a choice for the FI terms, as discussed in section \ref{subsec:abc}. The fact that the symmetry is recovered for the pure-Higgs states, seems to be additional evidence that these states are associated with the scaling point, and belong equally to the three different branches of the vacuum manifold obtained by the different choices of FI terms.

The generating function (\ref{ZB}) is made up of two non-trivial combinatorial\footnote{A generating function is called combinatorial if all its coefficients are positive.} factors: $Z_\beta=x^2y^2z^2 Z_\Delta Z_D$ with
\bea
Z_\Delta \equiv \frac{1}{(1-x y) (1-x z) (1-y z)}\qquad\qquad Z_D \equiv \frac{1}{(1-xy-yz-zx-2 x y z)}
\eea
As we show in appendix \ref{ap:comb}, the coefficients of the first factor can intuitively be thought of as a `delta function on even triangles':
\be
\Delta(a,b,c)=\begin{cases}1&\mbox{when }a+b+c \mbox{ even, } a+b\geq c\,,\ b+c\geq a \mbox{ and }c+a\geq b\\ 0&\mbox{otherwise}\end{cases}
\ee
The second factor $Z_D$ is a well known generating function in combinatorics \cite{Goulden}, its coefficients count the number of 3-derangements. More precisely $D(a,b,c)$ is the number of derangements, i.e. permutations without fixed points, of the multi-set that contains $a$ times $1$, $b$ times $2$ and $c$ times $3$. As is shown in appendix \ref{ap:comb}, from this combinatorial interpretation it follows that these coefficients vanish if and only if the triangle inequalities are violated:
\be
D(a,b,c)\neq 0\Leftrightarrow a+b\geq c\,,\ b+c\geq a \mbox{ and }c+a\geq b
\ee

Now note that the pure-Higgs degeneracies $\beta$ are simple convolutions of these combinatoric numbers:
\be
\beta(a,b,c)=\sum_{m,n,p=0}^{a-2,b-2,c-2}D(n,m,p)\Delta(a-n-2,b-m-2,c-p-2)
\ee 
First of all it follows from this formula that apart from Coulomb-states, additional pure-Higgs states are present on the Higgs branch if and only 
if the shifted intersection products $(a-2)$, $(b-2)$ and $(c-2)$ satisfy the triangle inequalities:
\be
\beta(a,b,c)\neq 0\quad \Leftrightarrow\quad a+b-2\geq c\geq 2\,,\ b+c-2\geq a\geq 2\,,\mbox{ and } c+a-2\geq b\geq 2\,. 
\ee
Finally it also gives us a first, somewhat involved combinatorial interpretation of these degeneracies. It turns out that $\beta(a,b,c)$ counts all derangements of $(a-2-m)$ numbers $1$, $(b-2-n)$ numbers $2$ and $(c-2-p)$ numbers $3$ for all $n,m$ and $p$ with an even sum and satisfying the triangle inequalities. It would be most interesting to rederive this combinatorial result from a set of simple physical principles, something which we leave as a problem for future work. 

\subsection{Growth of the pure-Higgs states}
In the previous subsection we gave the general characterization of when pure-Higgs states are present. Here we will give an exact, rather simple expression for their number, in the limit of large charges. A first such asymptotic analysis (of the Euler characteristic) when $a=b=c\gg1$ was made in \cite{Denef2007b}, by using the well-studied asymptotic behavior of Legendre polynomials. Since we have calculated the explicit partition function for the number of pure-Higgs states (\ref{ZB}) we can use some mathematical results on the asymptotics of meromorphic generating functions to extract the large-charge behavior. Our result reproduces that of \cite{Denef2007b} and generalizes it to cover the whole cone of possible $a,b,c$, as long as they are of the same order.

Generating functions are powerful tools in analyzing sequences. For example, the asymptotics of the sequence are encoded in the behavior of the generating function near its poles. We will use the mathematical framework developed in \cite{Wilson2002,Wilson2005}, that  makes this relation precise for meromorphic generating functions of multiple variables. Consider a multivariable sequence $a_r$, $r=(r_1,\ldots,r_d)$ and its generating function $Z(x)=\sum_r a_r x^r$, with $x^r=(x_1^{r_1},\ldots,x_d^{r_d})$. If we can write $Z=\frac{G}{H}$, with $G$ non-vanishing and holomorphic near a smooth, strictly minimal\footnote{For the definition of what the condition to be strictly minimal implies see  \cite{Wilson2002,Wilson2005}.}, simple zero $x_\star$ of the holomorphic function $H$, then \cite{Wilson2002,Wilson2005} show that as $|r|\rightarrow\infty$ the sequence has the asymptotics\footnote{We rewrote the result of theorem 3.5 of \cite{Wilson2002} in a coordinate covariant form, as detailed in appendix \ref{ap:cov}.}
\be
a_{r}\sim \left(\frac{N(r)}{2\pi}\right)^\frac{d-1}{2}\left.\frac{G}{\sqrt{\left(\det DDH\right) (DDH)^{ij}D_iHD_jH}}\right|_{x=x_\star(r)} x_\star(r)^{-r}\label{asform}
\ee
An important role is played by the differential operator $D_i=\frac{\partial}{\partial\log x_i}$. It is for example used in the definition of the 'Hessian' matrix $(DDH)_{ij}=D_iD_jH$ and its inverse $(DDH)^{ij}=(DDH)^{-1}_{ij}$. The relation between the direction of $r$ and the corresponding pole $x_\star(r)$, together with the normalization factor $N(r)$, are found as the solution to the algebraic equations:
\be
D_iH(x_\star(r))=-N(r)r_i\quad\mbox{and}\quad H(x_\star(r))=0\label{poles}
\ee

Let us now apply this technology to the generating function of pure-Higgs states (\ref{ZB}):
\be
Z_\beta=\frac{x^2y^2z^2}{(1-xy)(1-yz)(1-zx)(1-xy-yz-zx-2xyz)}
\ee
We can take $G=\frac{x^2y^2z^2}{(1-xy)(1-yz)(1-zx)}$ and $H=1-xy-yz-zx-2xyz$. The relation between the zeros of $H$ and the asymptotic direction in $abc$-space can then be found by solving (\ref{poles}):
\bea
x_\star&=&\frac{B C}{2a A}\\
y_\star&=&\frac{C A}{2b B}\label{ourpoles}\\
z_\star&=&\frac{A B}{2 c C}\\
N&=&\frac{aA+bB+cC}{4abc}
\eea
where for convenience we introduced the three `triangle functions' 
\be
A\equiv-a+b+c,\quad B\equiv a-b+c,\quad C \equiv a+b-c
\ee
Note first that the conditions for the poles to lie in the first quadrant are equivalent to the triangle inequalities for $a,b,c$: $A\geq0,B\geq0$, $C\geq0$. This is in perfect agreement with the fact that only when these inequalities are satisfied a non-zero degeneracy exists, as we showed in the previous subsection. This also justifies our assumption of analyticity of $G$ near the zeros of $H$:
\be
G(x_\star,y_\star,z_\star)=\frac{A^2B^2C^2}{\left(aA+bB+cC\right)^3}
\ee
What remains is to compute the nontrivial factor
\bea
\left.\left(\det DDH\right) (DDH)^{ij}D_iHD_jH\right|_{x_\star}
&=&2x_\star^2y_\star^2z_\star^2\left(y_\star z_\star(2x_\star^2+2x_\star+2+ y_\star+z_\star) \right. \nonumber \\
 +z_\star x_\star(2y_\star^2+2y_\star+2 &&\!\!\!\!\!\!\!\!+z_\star+x_\star)
+ \left. x_\star y_\star(2z_\star^2+2z_\star+2+x_\star+y_\star)\right)\nonumber\\
&=&\frac{ABC(aA+b B+cC)}{256\,a^3b^3c^3}
\eea
One now has all the ingredients to put together formula (\ref{asform}) to find the number of large-charge pure-Higgs states. For $a,b,c\gg1$
\be
\beta(a,b,c)\sim\frac{2}{\pi}\sqrt{\frac{abc(ABC)^3}{(aA+bB+cC)^7}}\frac{a^ab^bc^c}{A^AB^BC^C}\,2^{a+b+c}
\ee
and hence the number of pure-Higgs states grows exponentially with the charges when all intersection products are large and satisfy the triangle inequalities. This can be shown explicitly: First note that
\be
\beta(\l a,\l b,\l c)\sim\frac{2}{\pi}\sqrt{\frac{abc(ABC)^3}{(aA+bB+cC)^7}}\frac{1}{\l}\left(\frac{(2a)^a(2b)^b(2c)^c}{A^AB^BC^C}\right)^\l
\ee
Since $(A+B)(A+C)\geq A^2$ and cyclic, it follows that
\bea a \log2a\,+b\log2b\,+c\log2c&=&\frac{A}{2}\log(A+C)(A+B)\,+\frac{B}{2}\log(B+C)(A+B)\nonumber\\&&+\frac{C}{2}\log(A+C)(C+B)\nonumber\\&\geq& A\log A+B\log B+C\log C\nonumber
\eea
This implies that $\frac{(2a)^a (2b)^b (2c)^c}{A^A B^B C^C}\geq 1$ in the scaling regime and so $\beta$ grows exponentially in the scale $\l$.

Finally in the limit of equal intersection products, $a=b=c$, we reproduce the result of \cite{Denef2007b}:
\be
\beta(a,a,a)\sim\frac{2}{3^{7/2}\pi}\frac{2^{3a}}{a}\,.
\ee

\section{The Higgs-Coulomb Map and a Decoupling Limit}\label{sec:hc}

Let us consider again the Lagrangian for the quiver quantum mechanics (\ref{eqn:lagrangian}). In this section it will be convenient to switch to a canonical field theory convention in which the kinetic terms appear as
\begin{equation}\label{eqn:normlag}
{\cal L}={1\over g_{YM}^2} ({\dot X}^2 + D^2+2i{\bar\lambda}{\dot\lambda}) + |{\dot \phi}|^2+...
\end{equation}
In these convention $[g_{YM}^2]=3,\ ([X]=1,[\lambda]=3/2,[D]=2),\ ([\phi]=-1/2,[\psi]=0),\ [\theta]=-1$ . 

This Lagrangian has two distinct IR limits $g_{YM}^2\rightarrow\infty$, distinguished by the way in which the fields are rescaled as the IR limit is taken (see \cite{Witten:1997yu,Aharony:1999dw} for the 1+1 dimensional case). The two limits are:
\begin{itemize}
\item{\bf IR Coulomb-branch limit:} In this limit ${\hat X}=X/g_{YM}$ is kept fixed as $g_{YM}^2\rightarrow\infty$. This IR limit truncates the Hilbert to states that are asymptotically far on the Coulomb branch. On such one obtains a flat metric for ${\hat X}$, with small corrections that come from integrating out the chiral multiplets at a mass scale ${\hat X}g_{YM}$. 
\item{\bf IR Higgs-branch limit} In this limit all fields are held fixed as $g_{YM}^2\rightarrow \infty$. The kinetic term for the vector multiplet fields goes to zero, and they become auxiliary variables. The only dynamical fields are the chiral multiplets. After solving for the auxiliary vector fields in terms of the chiral fields we obtain a non-linear sigma model on the Higgs branch.
\end{itemize}
The way that we scale the FI parameter in the different limits is more subtle and we discuss it in appendix \ref{app:decoupling}.

Since $X$ is scaled radically differently, the Higgs branch is disconnected from the wave functions captured by the first limit. In the Higgs-branch limit, which is our main focus, the vector multiplet fields are simply specific operators on the Higgs branch. To the extent that the dynamics of some of the states on the Higgs branch can be described in terms of the dynamics of these operators, we can think of these states as moving on a Coulomb branch-like throat that emanates from the Higgs branch, and we can refer to this loosely as a Higgs-Coulomb equivalence. These are exactly the non-middle-cohomology states described above.

%
%As we have seen in the previous section we have a rather detailed understanding
%of the structure of the Higgs branch and how states of the latter map
%(subjectively) onto states of the Coulomb branch.  At $g_s =0$ the
%supersymmetric sector can be described entirely in terms of the Higgs branch
%but, at finite $g_s$ and when the charges are scaling, this branch is
%continuously connected to the Higgs branch and supersymmetric states correspond
%to wavefunctions spreading over both branches.  Such wavefunctions, in
%principle, encode the spacetime structure of the generic states of the
%partition function (\ref{ZE}) but solving the quantum mechanics of the combined
%branches is rather daunting.

The simplest context in which this was carried out in detail, in quantum mechanics, is the sigma model on the ADHM moduli space \cite{Berkooz1999}\footnote{A discussion of the two branches in 1+1 dimensional (0,4) and (4,4) GLSM models appears in \cite{Witten:1997yu,Aharony:1999dw} and in a partition function in \cite{Dorey:1999pd}.}. As an aside, the black holes that we are discussing here can be enumerated using the (0,4) CFT of \cite{Maldacena1997}. Note that the latter is not obtained in any simple way from a linear sigma model, which is our starting point here. Nevertheless, there are singular points in the the moduli space of M5 branes on a CY in which it the M5 brane can decompose into several ones, where a Higgs-Coulomb equivalence of the type that we are discussing might exist. The basic objects that we are using to build the Higgs-Coulomb equivalence - the SU(2) symmetry generators via the Lefschetz action - are also there in the MSW model, where they are part of an $SU(2)_R$ current algebra.  Constructing a Higgs-Coulomb equivalence for the MSW string would be a generalization of the D1-D5 system to lower SUSY, just as the quiver that we are discussing here is a generalization to lower SUSY of the D0-D4 system.

After taking $g_{YM}^2\rightarrow\infty$, one obtains the following equations from varying the Lagrangian with respect to the vector multiplet auxiliary fields
\begin{equation}\label{eqn:deceom}
	\theta_p +  \sum_{q\rightarrow p} s_{pq} |\phi_{pq}|^2 = 0, \qquad 
	\sum_{q\rightarrow p} 2  x^i_{pq} |\phi_{pq}|^2 +  s_{pq}\bar{\psi}_{pq}\sigma^i \psi_{pq} = 0, \qquad
	\sum_{q\rightarrow p}  s_{pq}\bar{\phi}_{pq} \epsilon \psi_{pq} = 0\,.
\end{equation}
Moreover in this limit $D$ becomes a Lagrangian multiplier exactly enforcing the
D-term constraint.  For more details of this limit see appendix
\ref{app:decoupling}. We must be careful, however, in taking the limit of \cite{Berkooz1999} as it is
essentially a decoupling limit and from \cite{Boer2008b} it is clear that such
limits may cause multicenter solutions to decay.  In Appendix
\ref{app:decoupling} we show that it is possible to take this limit in a way
that preserves some multicentered states and indeed this corresponds precisely
to the AdS$_3\times$S$^2$ decoupling limit of \cite{Boer2008b}.   An important
consequence of this limit is that $\theta_p$ in (\ref{eqn:deceom}) is non-zero
iff the center has D6 charge ($p_p^0 \neq 0$).

The key point is that in this limit the field that usually parameterize the Coulomb-branch, $x^i_{pq}$, is not set to zero; rather its equation of motion implies that, schematically,
\begin{equation}\label{eqn:higgscoulomb1}
	x_{pq}^i = \frac{\bar\psi_{pq}\sigma^i \psi_{pq}}{2 |\phi_{pq}|^2}
\end{equation}
which can now be interpreted as an operator relation defining an operator
$\hat{x}_{pq}$ on the Higgs branch.  We will see that the VEVs of this new
operator parameterize a space that is essentially the Coulomb branch.  Thus, in
this limit, the Coulomb branch emerges from a change of basis in the Higgs
branch \cite{Berkooz1999}.  In fact (\ref{eqn:higgscoulomb1}) is precisely the
change of basis that maps the Lefchetz action in the Higgs branch to the
angular momentum operator in spacetime.  

As we explain below, the full story is somewhat more complicated; in particular Eq. 
(\ref{eqn:higgscoulomb1}) is somewhat non-trivial to derive for more than two
centers. Technicalities aside, however, the two-center problem appears to capture
the essential physics of the map so let us begin by reviewing it.

\subsection{Two Centers}

For two centers $pq = 12$, the sum in the first equation of
(\ref{eqn:deceom}) has only one term and therefore (\ref{eqn:higgscoulomb1}) holds
immediately.  This operator relation becomes much more natural if we rephrase
it in terms of the angular momentum operator on the Coulomb branch.  Recall
from \cite{Berglund2006,Boer2009} that the latter is given by
\begin{equation}
	\hat{J}_c^i  = \frac{1}{2}\sum_{p < q} \frac{\Gamma_{pq} \, x^i_{pq}}{|x_{pq}|}
\end{equation}
with the subscript $c$ emphasizing that this is a Coulomb branch angular
momentum.  For two centers equations (\ref{eqn:int}) and (\ref{eqn:htheta}) imply
that $x_{12} = \Gamma_{12}/2\theta_1$ so
\begin{equation}
	\hat{J}_c^i = \theta_1 \,x_{12}^i = - \theta_1\, s_{12} \frac{\bar{\psi}\sigma^i \psi}{2 |\phi|^2} = \frac{1}{2} \bar{\psi}\sigma^i\psi
\end{equation}
where we have dropped the $pq$ labels on the hypermultiplets (we only have two centers) and in the second and third equality we have imposed the $D_p$ and $x_p^i$ e.o.m. (the first and second equation of (\ref{eqn:deceom})).

To understand the nature of this operator recall that $\psi_A, \bar{\psi}^A$
($A=1,2$) are two-component fermions with non-vanishing commutation relations
\footnote{See \cite[Appendix A]{Denef2002} for fermion conventions. In
\cite{Denef2002} Greek letters $\alpha, \beta$  are used for fermionic indices
whereas here they denote flavor indices $\alpha=1,\cdots,\Gamma_{pq}$ and we use upper-case Latin characters for fermionic indices.}
\begin{equation}
\{\bar{\psi}^{1\, \a}, \psi^\b_1\} = \delta^{\a\b}, \qquad \{\bar{\psi}^{2 \,\a}, \psi^\b_2\} = \delta^{\a\b}\, .
\end{equation}
These commutation relations and the supersymmetry variations in
\cite{Denef2002} are consistent with the identification of $\psi$ as
(holomorphic) differentials and derivatives on the target space of the theory 
\begin{equation}
	\bar{\psi}^{1\baa} \rightarrow d\bar{\phi}^\baa, \qquad
	\psi^\a_1 \rightarrow g^{\a\bbb} \frac{\partial}{\partial\, d\bar{\phi}^\bbb}, \qquad
	\bar{\psi}^{2\baa} \rightarrow g^{\baa\b} \frac{\partial}{\partial\, d \phi^\b}, \qquad
	\psi^\a_2 \rightarrow d\phi^\a \, ,
\end{equation}
from which we see that $\hat{J}_c^i$ is nothing else but the Lefschetz action
on ${\mathbb C}^{\Gamma_{12}}$ (the target space of the $\phi^\a$)
\begin{eqnarray}
	\hat{J}_c^3 &=& \frac{1}{2} \left(d\bar{\phi}^\baa \wedge \frac{\partial}{\partial\, d\bar\phi^{\bar\a}} + d\phi^\a \wedge \frac{\partial}{\partial\,d\phi^\a} - \Gamma_{12}\right), \nonumber\\
\hat{J}_c^+ &=& g_{\baa \a}d\bar{\phi}^\baa \wedge d\phi^\a, \\
\hat{J}_c^- &=&g^{\baa \a} \frac{\partial^2\quad}{\partial\,d\bar\phi^\baa\partial\, d\phi^\a}\nonumber
\end{eqnarray}
with $\hat{J}_c^\pm$ adding/removing a power of the symplectic form and
$\hat{J}_c^3$ giving $(p + q- \Gamma_{12})/2$ when acting on an element of
$H^{p,q}({\mathbb C}^{\Gamma_{12}})$.  

We have not yet imposed the last e.o.m. in (\ref{eqn:deceom}), which comes from the variation with respect to 
$\lambda$.  This equation projects the $\psi$ to the tangent bundle of the
$\mathbb{CP}^{\Gamma_{12} -1}$ which comes from imposing the D-term constraints on
$\phi_{12}$
\begin{equation}
	-s_{12}	|\phi_{12}|^2 = \theta_1 \, .
\end{equation}
This space is the vacuum moduli space of the Higgs branch and this projection
will also pull back $\hat{J}_c$ above to give the Lefshetz action on the
cohomology of $\mathbb{CP}^{\Gamma_{12}-1}$.

Thus we see explicitly that the operator relation (\ref{eqn:higgscoulomb1})
maps the spacetime angular momentum operator (whose eigenvalues characterize
the two-center states) to the Higgs-branch Lefschetz operator.  The two-center Coulomb branch vacuum
manifold is an S$^2$ and
$\hat{J}_c^i$ is just the quantization of this sphere. From the point of view of this quantization the two-sphere re-emerges
in the classical, $\Gamma_{12} \rightarrow \infty$, limit.

This map between Higgs- and Coulomb-branch states extends to three centers, but, as we will see, its structure is far less trivial: the overall Lefschetz
action will still map to the total spacetime angular momentum, but we will also find
operators that measure the positions of individual centers.

\subsection{Three Centers}

In order to solve the $x$ equations of motion for three centers we first use translational invariance to fix $x_3=0$ and
then solve 
\begin{equation}
\left(
\begin{array}{ccc}
	\phi _{12}^2+\phi _{31}^2 & -\phi _{12}^2 & -\phi_{31}^2 \\
 -\phi _{12}^2 & \phi _{12}^2+\phi _{23}^2 & -\phi _{23}^2 \\
 -\phi _{31}^2 & -\phi _{23}^2 & \phi _{23}^2+\phi _{31}^2
\end{array}
\right)
\left(
\begin{array}{c}
 x^i_1 \\
 x^i_2 \\
 0
\end{array}
\right) 
= 
-
\left(
\begin{array}{c}
 s_{12} b^i{}_{12} -s_{31} b^i{}_{31}  \\
 s_{23} b^i{}_{23} -s_{12} b^i{}_{12}  \\
 s_{31} b^i{}_{31} -s_{23} b^i{}_{23} 
\end{array}
\right)
\end{equation}
with 
\begin{equation}
	b_{pq}^i \equiv \frac{\bar{\psi}_{pq} \sigma^i \psi_{pq}}{2} \, .
\end{equation}
The solution for $x^i_1$, $x^i_2$ is
\begin{equation}
	\begin{split}
x_1^i = -\frac{s_{23} \phi _{12}^2 b^i{}_{23}-s_{31} \phi _{12}^2 b^i{}_{31}+s_{12} \phi _{23}^2 b^i{}_{12}-s_{31} \phi _{23}^2 b^i{}_{31}}{\phi _{12}^2 \phi _{23}^2+\phi _{31}^2 \phi _{23}^2+\phi _{12}^2 \phi _{31}^2} \\
x_2^i = -\frac{s_{23} \phi _{12}^2 b^i{}_{23}-s_{31} \phi _{12}^2 b^i{}_{31}-s_{12} \phi _{31}^2 b^i{}_{12}+s_{23} \phi _{31}^2 b^i{}_{23}}{\phi _{12}^2 \phi _{23}^2+\phi _{31}^2 \phi _{23}^2+\phi _{12}^2 \phi _{31}^2} \,.
	\end{split}
\end{equation}
While this expression is clearly much more complicated than for two centers,
we will see it simplifies significantly when the quiver is closed
and there is a superpotential.  

As we have explained in Section \ref{subsec:abc}, when
the quiver is closed the superpotential $W(\phi) \neq 0$ and moreover is
effectively cubic \cite{Denef2007b}.  Generic solutions to this superpotential then must have
$\phi^\a_{pq} = 0$ for one of the hypermultiplets; the choice of which
hypermultiplet vanishes is dictated by the sign of the FI term and, in keeping
with section, \ref{subsec:abc} we will take  $\phi_{31} = 0$.  Moreover, since in
a closed quiver the graph is directed the $s_{pq}$ have the same sign, which we take $s_{12} = s_{23} = s_{31} = -1$.

This yields
\begin{equation}
	x_{12}^i = \frac{b^i{}_{12}}{\phi _{12}^2}, \qquad
	x_{23}^i = \frac{b^i{}_{23}}{\phi _{23}^2}, \qquad
	x_{31}^i = -\frac{b^i{}_{12}}{\phi _{12}^2}-\frac{b^i{}_{23}}{\phi _{23}^2}\, .
\end{equation}
For this choice the D-term conditions also reduce to:
\begin{equation}
	\phi_{12}^2 = \theta_1, \qquad \phi_{23}^2 = - \theta_3
\end{equation}
and hence the spacetime positions reduce to simple operators 
\begin{equation}\label{eqn:xmap}
	x_1^i = \frac{\bar{\psi}_{12} \sigma^i \psi_{12}}{2 \theta_1} -
	\frac{\bar{\psi}_{23} \sigma^i \psi_{23}}{2 \theta_3}, \qquad
	x_2^i = - \frac{\bar{\psi}_{23} \sigma^i \psi_{23}}{2 \theta_3}
\end{equation}
and the total angular momentum is just the Lefschetz operator on on ${\mathbb
C}^a \times {\mathbb C}^b$
\begin{equation}\label{eqn:jmap}
	\hat{J}_c^i = \frac{1}{2} \left(
	\frac{\Gamma_{12}}{x_{12}} x^i_{12} +
	\frac{\Gamma_{23}}{x_{23}} x^i_{23} +
	\frac{\Gamma_{31}}{x_{31}} x^i_{31} \right) = 
\frac{\bar{\psi}_{12} \sigma^i \psi_{12}}{2} +
	\frac{\bar{\psi}_{23} \sigma^i \psi_{23}}{2} 
\end{equation}
where we've used the fact that $x_{31}^i = - x_{23}^i - x_{12}^i$ and also used the
constraints (\ref{eqn:int}) combined with (\ref{eqn:htheta}) to eliminate the
$\theta$'s.

We see therefore that for three centers the Coulomb branch is only non-trivially reproduced once we've imposed both the D- and F-term constraints.  Moreover, as is clear from the discussion in section \ref{sec:cohomology} the ``pure Higgs'' states all map to zero-angular-momentum states on the Coulomb branch.  While it is not evident from (\ref{eqn:xmap})-(\ref{eqn:jmap}) we know from an independent analysis of the three-center Coulomb branch \cite{Bena:2007qc,Boer2009} that the only point in the three-center solution space with zero angular momentum is the scaling point when all $\vec{x}_p = 0$.  This is the only way to set (\ref{eqn:jmap}) to zero while satisfying the constraints (\ref{eqn:int}).

\subsection{Some comments on Higgs branch corrections to the multi-center dynamics}

 In the limit that the centers are close to each other, gravity develops a long $AdS_2$ throat. The associated conformal symmetry is manifested in the effective Lagrangian of the centers. We would like to see if can say something about the reliability of this throat from the construction above.

First of all, it is worth noting that the moduli space is not entirely self-consistent, as it contains trajectories that pass through the origin. To see this consider a trajectory that goes through the origin, say along the $x^1$ direction. The classical trajectory is: $x^1(t)=x^1_0/t^2$. The effective Lagrangian is an expansion in velocities. Any term in the conformally-invariant Lagrangian has the same time dependence as $(1/x)\partial_t$ to some power times the leading term. However, $(1/x)\partial_t$ scales as $t$ and since near the origin $t\rightarrow \infty$ all these terms become large. Higher-order terms, suppressed by a mass term $M$, would be less important because are they are suppressed by $(1/M)\partial_t\sim 1/t$ or $x/M\sim 1/t^2$ (but with a high enough power of $(1/x)\partial_t$ they may still appear in a diverging term).

Computing such terms, however, is complicated - it requires a GR computation of the effective action beyond the moduli space approximation, or with higher order terms. In the absence of these computations, one can use the physics of the Higgs branch and the Higgs-Coulomb map to understand some the effects that will cut off the throat. There are two sources for such corrections:

1. The $X$ are not real bosonic variables - rather they are fermion bilinears. The dynamics in the effective action cannot be trusted when the wavelength of the quantum wave packet in the $X={{\bar\psi}\sigma\psi}/\phi^2$ goes below $1/\theta$. In this case, if we wrote the full dynamics, we would see the quantization or granularity of X as a fermion bi-linear. In the effective action this in the term which is linear in the time derivative, which was used in \cite{Boer2009} to argue that the throat should be cut off a minimal value of the angular momentum and $x$. The identification of the position as a composite fermion operators indicates that this is true even beyond the low energy limit.

2. Recall that we have been using a linear sigma model description of the Higgs branch, obtained from the quiver quantum mechanics. The latter describes only a part of the Higgs branch, where the VEV's of the fields are small enough. The full ``Higgs branch'' is presumably the moduli space of branes wrapping cycles in the CY, with the appropriate charges. This conjecture is natural given the MSW counting of the entropy of these black holes. The linear sigma model that we used is therefore only an effective model in a specific regime of the full moduli space, and we expect that it will be corrected by higher-order terms. 

Since the linear sigma model is effective for small values of the fields (when the centers are close to each other and the chiral multiplets are light), the information about the rest of the moduli space is captured when the values of the chiral fields are large. This implies that there exists a scale $\mu$, with dimension $-1/2$ such that when the chiral fields $\phi\sim\mu$ the dynamics of the Higgs branch deviates from the quiver quantum mechanics expectation. In particular, when $X$ is smaller than $1/\mu^2$, the field $\phi$ can spread all the away to $\phi\sim\mu$ and the dynamics gets corrected. Roughly we expect that the dependence on $X$ in the effective Coulomb branch dynamics will be replaced by a dependence on $X+1/\mu^2$, changing the dynamics at small $X$. This point can be made more precise for the D0-D4 system, to which we turn in the next subsection

\subsection{The Higgs-Coulomb equivalence for a compact D0-D4 system}

As we mentioned before, the D0-D4 system is a good testing ground for the ideas discussed in this paper.
%, in particular on the Higgs-Coulomb map and its validity.
The theory has 8 supercharges, and hence we expect that by analyzing it we will also be able to shed some light directly on the conformal symmetry found in \cite{Michelson:1999dx}, and, more interestingly, on its violation in the throat.

%\begin{equation} \label{aux2}
{%1\over g_sl_s} \int dt \biggl( (\partial X)^2 + \rho\partial \rho + (\partial Q)^2 + %\psi\partial\psi + {1\over l_s^4}X^2QQ + X\psi\psi+{1\over l_2}^4 (QQ)^4+...\biggr)
%\end{equation}
%X are 0-0 strings ($\rho$ is their superpartner), Q are 0-4 strings ($\psi$). The term before last %are the D and F terms (written very schematically) and the ... are some couplings of $\rho,\psi$. %The $X^2QQ$ and $X\psi\psi$ are couplings of the vector multiplet to the hypermultiplet (for %example by dimensional reduction from 5+1) and the $(QQ)^2$ are the F and D-terms. The %system has 8 supercharges under which the Q's are a hypermultiplet and the X's are a vector %multiplet (alongside with the $U(1)$ gauge field, $A_0$, on the D0-brane which is not written %above). The dimensions of the fields are
%\begin{equation}
%[X],[Q]=-1,\ [\rho],[\psi]= -1/2,\ [A_0]=1.
%\end{equation}
%For simplicity we can consider a single D0-brane and K D4-branes.

In the language of four-dimensional ${\cal N}=2$ multiplets, the low-energy theory of $N_0$ D0 branes and $N_4$ (compact) D4 branes has a vector multiplet $(X,\rho,D)$ in the adjoint of $SU(N_0)$, and hypermultiplets $(Q,\psi)$ charged in the fundamental of both $SU(N_0)$ and the global $SU(N_4)$ symmetry.  
The metric on the D0 moduli space (for a single brane, for simplicity), can be obtained by integrating out the D0-D4 string, and is \cite{Douglas:1996yp}:
\begin{equation} \label{aux3}
\biggl(1+{g_s N_4 l_s^3\over r^3}\biggr)(\partial X)^2
\end{equation}
where $r^2=\Sigma X^2$. This is a one-loop result, but if one assumes SO(5) symmetry, and some other mild assumptions one can show that the metric is actually not renormalized further \cite{Diaconescu:1997ut}. As usual, to go to the near-horizon region we  ``drop the 1,'' and this opens up a new non compact space  at $r=0$, with the same scaling symmetry as in the black holes quivers discussed above
\begin{equation}
t\rightarrow\lambda t,\ X,\rightarrow \lambda^{-1}X,\ \rho\rightarrow \lambda^{-3/2}\rho,\ D\rightarrow \lambda^{-2}D
\end{equation}

When the D4 brane is non-compact this conformal symmetry is exact, and not just an artifact of the moduli space approximation. Indeed, taking the limit $g_{YM}^2$ which puts the system on the Higgs branch is the same as neglecting the $1$ in the Coulomb branch metric. The kinetic terms for the vector multiplet vanish in this limit, and it is easy to see that the when $\theta=0$) this theory has a scaling symmetry:
% , when also adding the transformation law 
\begin{equation}
Q\rightarrow \lambda^{1/2}Q,\ \psi\rightarrow \psi,\ F\rightarrow \lambda^{-1/2}F 
\end{equation}
This is, of course, compatible with the relation $X={\bar \psi}\psi/Q^2$.

If we have a $\theta$ term in the Lagrangian, it becomes irrelevant when one is in the throat, and disappears as $\lambda\rightarrow 0$. As argued above this is similar to the scaling limit, in which the $\theta$ also become irrelevant in the bubble equations as the $x_{ij}$'s are going to zero. The similarity is not a coincidence - we will shortly put the D0-D4 on a 6 torus to obtain a 4D model.

The map $X=\bar\psi\psi/Q^2$ has the following qualitative interpretation: Within the moduli space of instantons there are singular points in which the gauge symmetry is enhanced. Physically, at these points some of the instantons shrink. When an instanton shrinks inside a D4 brane, it can leave the D4 brane as a D0 brane, which implies that the Coulomb branch is attached to the Higgs branch at the shrinking point. For example, when $N_0=1$ and $ N_4=2$ the Higgs branch is $\mathbb{R}^4_{\mathrm{center of mass}}\times \mathbb{R}^4/Z_2$. The near-Higgs Coulomb branch is a description of the Higgs branch degrees of freedom near these singularities, and since ${\bar \psi}\psi$ is quantized, then $X\sim 1/Q^2$. The smaller $Q$ is and the closer we are to the singularity, the more strongly coupled the description in terms of the Higgs branch variables becomes. On the other hand, the Coulomb-branch description in terms of $X$ improves as $X$ becomes large. Conversely, when $X$ becomes small we are pushed to larger values of $Q$. In this regime the Higgs-branch variables are weakly coupled, whereas the $X$ variables are strongly coupled, indicating the limit of validity of ``Coulomb-branch'' description in terms of $X$ at small values in the full action.

When the D4-brane is compact, the moduli space of instantons is compact as well. We will denote the size of the radii of the $T^4$ by $L$ (all radii are roughly the same). The 1+1 dimensional analogue of this quantum mechanics is the D1-D5 system, which is a sigma model on a deformation of $(T^4)^{N_0N_4}/S_{N_0N_4}$.

Quantum Mechanics on this space has no obvious conformal symmetry. However, the points at which the manifold develops a singularity are locally the same, because they are associated with zero size instantons, for which the global structure of the $T^4$ is irrelevant.  This is similar to the statement that the local singularities in $T^4/Z_2$ are the same as in $R^4/Z_2$. There is therefore an approximate conformal symmetry in the vicinity of these points. The symmetry is broken at large value of $Q$ and becomes more exact at small values of $Q$. Translating to the Coulomb branch variables, the conformal symmetry is a better approximation at large values of $X$, and breaks down for small values of $X$. It is easy to evaluate the scale where this happens: If the size of the compact manifold of the $Q$'s is characterized by a scale, $\zeta$, which has dimension $-1/2$, then the cut-off in the Coulomb-branch description is at $X<\zeta^{-2}$. 

One can write a model which takes these effects into account, but one can also obtain it using scaling arguments. The corrections to the effective action should disappear in the limit $\zeta\rightarrow\infty$ which means that they give an expansion in $1/\zeta$. Compared to the conformal terms, the terms in this expansion would have additional powers of $1/(X\zeta^2)$. Hence, these terms blow up at $X\rightarrow 0$.  

{%g_s\over M_sL^2} (\partial X)^4/X_0^8
%\end{equation}
%(this is all dimensional analysis if the diagram does not cancel from SUSY. In a conformal theory %$(\partial X)^4$ will appear as $(\partial X)^4\over X^7$ and the new vertex has to come as %${1\over X_0 L^2M_p}$).
%
%
%2. Approach 2 requires integrating out Q when its restricted to be in a box of size L. I have only %bits and pieces of it so I will add it in later.

A related situation happens in the conformal quantum mechanics for five-dimensional black holes, described in \cite{Michelson:1999dx}. The dynamics on the moduli space is a conformal quantum mechanics, with 8 supercharges, but with a slightly different kinetic term. When the centers approach each other the kinetic term behaves like
\begin{equation}
{\cal L}\sim {{(\dot U)}^2\over U^4}\, .
\end{equation}       
The conformal symmetry here is not of the same form as the scaling symmetry we encountered in four dimensions. Indeed, this model is described by intersecting M2 branes wrapping 2 cycles in the CY, for  which there is no quiver model because the theory on the M2 branes is not an ordinary gauge theory. The main branch of this moduli space is when all the M2 branes are connected, but there are points of degeneration where a single connected M2 can split into distinct components. The geometric part of the moduli space of M2 wrapping cycles in a CY is the same as the moduli space of D2 branes wrapping the CY (there could be additional moduli upon compactification) and is still compact. The latter is dual (for simple enough manifolds) to the D0-D4 moduli space, on another CY. Hence, we expect that the D0-D4 argument above can be applied to this system as well, indicating that the throat will be cut off at some small value of $X$. 

There is yet another way to argue that Coulomb branch is capped off. When the moduli space is compact there is a finite gap between the extremal states (encoded in the cohomology) and the first excited states. The gap goes to zero when the volume of the internal space becomes non-compact, and for the D0-D4 or the M2-M2 systems this volume will be determined by the string theory/M theory moduli. In black hole language, the absence of this gap at the quantum level means that the entropy of the black hole is infinite. For example, we can take $g_s$ to zero, keeping the volume of the compactification fixed in string units. This rescales the dimension (-1/2) $\zeta$ by $g_s^{-1/2}$ for, say, the D0-D4 strings. At the same time $M_p\rightarrow \infty$ and the entropy in any fixed-charge sector of the theory (which has a black hole) diverges. The conformal symmetry comes at a costly price - only when black holes in the theory have a continuum of states and infinite entropy can they accommodate a conformal symmetry. Any finite $M_p$ makes the black holes studied in \cite{Michelson:1999dx} have a finite number of states and hence they can no longer accommodate a conformal symmetry.

\subsection{How do the pure Higgs branch states look in supergravity?}

Given that the pure Higgs branch states live in the kernel of our Higgs-Coulomb map, it is interesting to ask how they look in the regime of parameters where supergravity is a valid description of the physics. Since they have zero angular momentum, and since the only three-center solution with zero angular momentum lives at the scaling point, one possibility is that the pure Higgs states will develop a horizon and will map to the single-center black hole. However, one can also build solutions that have zero angular momentum away from the scaling point\footnote{One example is the ``pincer'' solution of 
\cite{Bena2006d}.} and hence have a finite throat and no horizon; a Higgs state might also map into such a configuration. 

One can try to distinguish between the two possibilities by the following heuristic argument: Suppose we want to probe the pure Higgs states with gravity modes and find how much information can be obtained.  If one restricts for simplicity to modes of the metric, they couple to the action on the cluster of branes by terms prescribed by the DBI action. Suppose we label the position of the cluster of branes by $X=0$, where $X$ are the vector multiplet fields in the quiver. We can expand the gravity fields in derivatives around $X=0$, which gives an expansion in spherical harmonics around the point where the quiver sits, and see to which operators in the quiver they couple. In the near horizon limit when the coupling becomes large, this corresponds to probing a state on the Higgs branch with different modes of the gravitational field. 

Consider computing a 1-pt function of such fields from the quiver perspective. The time-time component of the metric, $g_{00}$, couples to the quiver energy which is determined by the BPS condition. On the other hand, the components $g_{0\mu}$ and any derivative of $g_{\mu\nu}$  couple to powers of $X^\mu$. The operator that we are evaluating, from the Higgs branch point of view, is therefore some power of $X^\mu$. 

There are now two possibilities: if one computes the expectation values of all such polynomials in a pure-Higgs cohomology state and some of them are nonzero this suggests that the state will correspond to a finite-size zero-angular-momentum configuration and hence will not have a horizon. If on the other hand all these operators have zero expectation values in the pure Higgs states, then the quadrupole, octopole, etc. moments of solutions corresponding to these states are all zero, and hence these states will be indistinguishable from a black hole, at least as far as gravity one-point functions are concerned. This calculation promises to shed light on this fascinating issue, and we leave it to future exploration.

\subsection*{Acknowledgements}
The authors are grateful for interesting and stimulating discussions with
F.~Denef, M.~Douglas, G.~Moore, P.~Paule, R.~Pemantle,
J.~Simon, and M.~Wilson.  They also would like to thank the
organizers of the ``The supersymmetric, the extremal and the ugly'' workshop at
CEA Saclay, where some of this work was presented and reanimated.  SE and DVdB
are grateful for the hospitality of the Simons Workshop in Mathematics and
Physics 2010 where some of this work took place. MB, JdB and SE would like to thank the Santa Barbara KITP for hospitality and support during the completion of this work via  the NSF grant PHY11-25915. The work of IB is supported in 
part by the ANR grant 08-JCJC-0001-0 and by the ERC Starting Independent Researcher 
Grant 240210 - String-QCD-BH. The work of MB is supported by the ISF center of excellence program, by the Minerva foundation, by the BSF and by GIF. The work of SE is supported primarily by the Netherlands Organization for Scientific Research (NWO) under a Rubicon grant and also partially by the ERC Starting Independent
Researcher Grant 240210 - String-QCD-BH. DVdB is supported by a TUBITAK 2216-research fellowship. This work is part of the research programme of the Foundation for Fundamental Research on Matter (FOM), which is part of the Netherlands Organization for Scientific Research (NWO).

\appendix

\newpage
\section{Two simple combinatorial computations}\label{ap:comb}
In this appendix we show two small technical results on the combinatorial coefficients appearing in the pure-Higgs generating function.
\subsection{$\Delta(a,b,c)$ as a delta-function}
We will compute the coefficients $\Delta(a,b,c)$ of the generating function
\be
Z_\Delta=\frac{1}{(1-xy)(1-yz)(1-zx)}
\ee
First note that
\be
\frac{1}{1-xy}=\sum_{a,b}\delta(a-b)x^ay^b
\ee
The coefficients $\Delta$ are thus simply a double convolution of such delta functions, one finds
\be
\Delta(a,b,c)=\sum_{m,p=0}^{a,c}\delta(m+p-b)\delta(a-m-c+p)
\ee
This is zero unless there exist $m,p$ such that $0\leq m\leq a$, $0\leq p\leq c$, $b=m+p$ and $a+b-c=2m$. One can check that this condition is equivalent to $a,b,c$ satisfying the triangle inequalities and $a+b+c$ even. In summary
\be
\Delta(a,b,c)=\begin{cases}1&\mbox{when }a+b+c \mbox{ even, } a+b\geq c\,,\ b+c\geq a \mbox{ and }c+a\geq b\\ 0&\mbox{otherwise}\end{cases}
\ee

\subsection{3-derangements and triangle inequalities}
Here we will show that the number of 3-derangements $D(a,b,c)$, i.e the number of permutations without fixed points of the multi-set that contains $a$ times $1$, $b$ times $2$ and $c$ times $3$, is non-vanishing if and only if $a,b$ and $c$ satisfy the triangle inequalities:
\be
D(a,b,c)\neq 0\Leftrightarrow a+b\geq c\,,\ b+c\geq a \mbox{ and }c+a\geq b
\ee
\begin{itemize}
\item First we show that if the three triangle inequalities are satisfied there always exists at least one derangement. By symmetry we can assume that $a\leq b\leq c$, and so the only non trivial inequality is $a+b\geq c$. On can then check then that the following permutation has no fixed points and is hence a derangement:
\be
(\underbrace{1,\ldots, 1}_{a\ \mathrm{times}},\underbrace{2,\ldots, 2}_{b\ \mathrm{times}},\underbrace{3,\ldots, 3}_{c\ \mathrm{times}})\mapsto
(\underbrace{3,\ldots, 3}_{c\ \mathrm{times}},\underbrace{1,\ldots, 1}_{a\ \mathrm{times}},\underbrace{2,\ldots, 2}_{b\ \mathrm{times}})
\ee
\item It is also easy to show that there are no derangements if the triangle inequalities are violated. Since derangements by definition don't allow fixed points we need to move all the $a$ numbers $1$ to another position, previously occupied by a $2$ or $3$. There are $b+c$ such positions, so this is only possible if $a\leq b+c$. By symmetry the other two triangle inqualities follow.
\end{itemize}

\section{A covariant formula for the asymptotics of multivariate sequences}\label{ap:cov}
In this appendix we document the algebraic manipulations that allow to write the result of Theorem 3.5 in \cite{Wilson2002} on the asymptotics of certain multivariate sequences in a covariant form.

In \cite{Wilson2002} the following formula for the asymptotics of a multivariate sequence with generating function of the form $Z=\frac{G}{H}$ is derived (under certain conditions):
\be
a_r\sim \left.(2\pi\, r_0)^{\frac{1-d}{2}}\frac{G}{D_0H\sqrt{\calh}} x^{-r}\right|_{x=x_\star(r)}\label{res1}
\ee
Here the following notation is implied $x=(x_0,\ldots,x_n)$, $r=(r_0,\ldots,r_n)$,  $n=d-1$. In this appendix we will use the following index notation: $i,j=0,\ldots, n$ and $a,b=1,\ldots n$. Furthermore
\bea
D_i=\frac{\partial}{\partial \log x_i}\,, \quad \calh=\det D_aD_b\log g\,,\qquad H(g(x_1,\ldots,z_{n}),x_1,\ldots,z_{n})=0
\eea
Note that the formula (\ref{res1}) should be evaluated at $x=x_\star(r)$, defined as the solution to the equations
\be
r_0D_iH(x_\star(r))=r_iD_0H(x_\star(r))\quad\mbox{and}\quad H(x_\star(r))=0\label{cond1}
\ee

The formula (\ref{res1}) is a beautiful and powerful mathematical result, but is in this form not manifestly coordinate invariant, as the coordinate $x_0$ plays a special role. We will show in this appendix how one can rewrite the formula in the following manifestly coordinate covariant form:
\be
a_r\sim \left.\left(\frac{N(r)}{2\pi}\right)^{\frac{d-1}{2}}\frac{G}{\sqrt{(\det DDH)(DDH)^{ij}D_iH D_jH}} x^{-r}\right|_{x=x_\star(r)}\label{res2}
\ee
In this formulation both $x_\star(r)$ and $N(r)$ are found by simultaniously solving the $(d+1)$ algebraic equations
\be
D_iH(x_\star(r))=-N(r)r_i\quad\mbox{and}\quad H(x_\star(r))=0\label{cond2}
\ee
Furthermore we used the compact notation 
\be
(DDH)^{ij}D_iH D_jH=\sum_{i,j=0}^n(DDH)^{-1}_{ij}D_iHD_jH
\ee

To relate the forms (\ref{res1}) and (\ref{res2}) first note that the conditions (\ref{cond1}) and (\ref{cond2}) are equivalent as we can take $D_0H=-N(r)r_0$ as the definition of $N$, furthermore note that also by definition $x_{\star\,0}=g$. The new form (\ref{res2}) then follows from the following identity
\be
\left.\det\left(DD\log g\right) \right|_{x_0=g}=\left.(-D_0H)^{-n-2}\left(\det DDH\right) (DDH)^{ij}D_iHD_jH\right|_{x_0=g}\label{baseid}
\ee
 
This identity can be obtained through some simple, but somewhat tedious algebra. A first step is to rewrite derivatives of $g$ in terms of derivatives of $H$. Note that since by definition $H(g,x_1,\ldots,x_n)$ is identically zero it follows that also $$D_{a_1}\ldots D_{a_k}H(g(x_1,\ldots,x_n),x_1,\ldots,x_n)=0\,.$$ Using this and the modified chain rule $D_i( f\circ h)=D_if\partial_ih$ one can derive that 
\bea
D_a g&=&\left.\frac{-D_aH}{\partial_0H}\right|_{x_0=g}\\
D_aD_bg&=&\left.\frac{-1}{\partial_0H}\left(D_aD_b H-\frac{D_0D_aHD_bH}{D_0H}-\frac{D_0D_bHD_aH}{D_0H}-\frac{D_aHD_bH}{D_0H}+\frac{D_0D_0HD_aHD_bH}{D_0HD_0H}\right)\right|_{x_0=g}\nonumber
\eea
Furthermore using that
\be
D_aD_b\log g=\left(\frac{1}{g}D_aD_bg-\frac{1}{g^2}D_agD_bg\right)
\ee
one finds that
\be
D_aD_b\log g=\left.\frac{-1}{D_0H}\left(D_aD_b H-\frac{D_0D_aHD_bH}{D_0H}-\frac{D_0D_bHD_aH}{D_0H}+\frac{D_0D_0HD_aHD_bH}{D_0HD_0H}\right)\right|_{x_0=g}\nonumber
\ee
The crucial step is to observe that this can be written as a product
\be
D_aD_b\log g=\frac{-1}{(D_0H)^3}V_{ai}D_iD_jH\,V^{\mathsf{T}}_{jb} \label{matrprod}
\ee
by introducing the $n\times(n+1)$ matrix
\be
V_{ab}=D_0H\delta_{ab}\,,\qquad V_{a0}=-D_aH\,.
\ee
On can now apply the Cauchy-Binet formula for the determinant of the product of non-square matrices to find
\be
\det(DD\log g)=(-D_0H)^{-3n}\sum_{i,j=0}^n\det V^{(i)}\det V^{(j)}\det(DDH^{(ij)})
\ee
Here $V^{(i)}$ is the $n\times n$ matrix obtained by removing the $i$'th column from the $n\times(n+1)$ matrix $V$, while $DDH^{(ij)}$ is the $n\times n$ matrix obtained by removing both the $i$'th row and $j$'th column from the $(n+1)\times(n+1)$ matrix $DDH$.
The identity (\ref{baseid}) then follows by observing that
\bea
\det V^{(i)}&=&(-1)^{i}\left(-D_0H\right)^{n-1}D_iH\\
\det(DDH^{(ij)})&=&(-1)^{i+j}\left({\mathrm{cofactor}}_{ij}DDH\right)=(-1)^{i+j}\left(\det DDH\right) (DDH)^{-1}_{ij}
\eea

\section{The Decoupling Limit}\label{app:decoupling}

The decoupling limit described in section 4 and \cite{Berkooz1999} sends $g^2_{YM}\rightarrow
\infty$ while simultaneously taking an IR limit of the theory.  We will do
something similar here but will phrase it in the language of \cite{Boer2008b} and Eq. (\ref{eqn:lagrangian}), 
so we can more carefully track its effect on the CY moduli (which were not so
important in \cite{Berkooz1999}).

The limit in \cite{Boer2008b} involves fixing the mass of stretched M2-branes
and hence we introduce $R$, the length of the M-theory circle, and lift all
quantities to 11-dimensions (see \cite[Appendix A]{Boer2008b} for conventions
and details).  In the decoupling limit we send, $\ell_{11}$, the Plank length in
11-dimensions to zero while fixing the  mass of M2's wrapping on $x^{11}$ and
the volume\footnote{Here we use $V_M$ for the volume measured by the 11-d
metric but this is not so important as we follow the conventions of
\cite{Boer2008b} where the IIA volume is set equal to the M-theory volume
asymptotically.} of the CY in Plank units
\begin{equation}
	M_{M2} \sim \frac{x R}{\ell_{11}^3}, \qquad \tilde{V}_M = \frac{V_M}{\ell_{11}^6}
\end{equation}
from which it follows that $J_M \sim \tilde{V}_M^{1/3}$, the Kahler moduli
normalized in Plank units, are fixed.  As in \cite{Boer2008b} we will take $R$
to be fixed in some arbitrary units implying that 
\begin{equation}
	x \sim \ell_{11}^3 \rightarrow 0, \qquad J_{IIA} \sim \left(\frac{R}{\ell_{11}}\right) J_M \rightarrow \infty
\end{equation}
where $J_{IIA}$ are the IIA moduli measuring volumes in string units, $J_{IIA}
\ell_s^2 = J_M \ell_{11}^2$ (from which the above follows via $R\ell_s^2 \sim
\ell_{11}^3$).  As explained in \cite{Boer2008b} this is a near-horizon limit that
also decompactifies to five-dimensions (as $R/\ell_{11} \rightarrow \infty$).

To understand how this limit affects the Higgs branch let us consider its
effects on $Z(\Gamma)$, the central charge associated with a center, which
appears in (\ref{eqn:mp})-(\ref{eqn:thetap}).  The $J$ in (\ref{eqn:mp}) is in
fact $J_{IIA}$ so schematically 
\begin{equation}\label{eqn:zscaling}
	Z(\Gamma) \sim 
	p^0 \left(\frac{R}{\ell_{11}}\right)^{3/2}  +
	q^A \left(\frac{R}{\ell_{11}}\right)^{1/2}  +
	q_A \left(\frac{R}{\ell_{11}}\right)^{-1/2} +
	q_0 \left(\frac{R}{\ell_{11}}\right)^{-3/2}  
\end{equation}
where we've only exhibited the scaling of each component of the charge.

As we are interested in taking an IR limit let us reintroduce factors of
$\ell_s$ in (\ref{eqn:lagrangian}) to correctly exhibit the dimensionality of
the couplings (for brevity we have dropped the center subscripts, $p,q$, but
one could equivalently think of this as the center-of-mass Lagrangian for two
centers
\cite{Denef2002})
\begin{equation*}\label{eqn:dimlag}
{\mathcal L} = \frac{m}{2} \left( \dot{x}^2 + D^2 + 2 i \bar\lambda \dot\lambda\right) - \frac{\theta D}{\ell_s} +
\frac{1}{\ell_s^2}\left[ \left(\frac{x^2}{\ell_s^2} + D\right) \phi^2 + \bar\psi \sigma^i x^i \psi  - i \sqrt{2}  ( \bar\phi \lambda \epsilon \psi - h.c. ) \right] 
\end{equation*}
Here we focus only on terms containing vector multiplet components as the other
terms will not play any role. Note that unlike in Section 4, $x, D$ and $\lambda$ have non-standard
dimensions due to their non-standard kinetic term, while $\theta$ is
dimensionless by (\ref{eqn:thetap}).

Let us now take the limit of \cite{Boer2008b} by rescaling all the components
of the vector multiplet by $\ell_{11}^{-3}$ (e.g. $\tx^i = x^i/\ell_{11}^3$ and
likewise for $\tD, \tL$)
\begin{equation*}\label{eqn:declag}
{\mathcal L} = \frac{m\ell_{11}^6}{2} \left( \dot{\tx}^2 + \tD^2 + 2 i \bar\tL \dot\tL\right) - \frac{\ell_{11}^3 \theta \tD}{\ell_s} +
R \left[ \left(R \tx^2 +  \tD\right) \phi^2 + \bar\psi \sigma^i \tx^i \psi  - i \, \sqrt{2}  ( \bar\phi \tL \epsilon \psi - h.c. ) \right] 
\end{equation*}
where we have used the fact that $R \ell_s^2 \sim \ell_{11}^2$.  Combining
(\ref{eqn:zscaling}) with (\ref{eqn:mp})-(\ref{eqn:thetap}) we find
\begin{equation}\label{eqn:mp2}
	m\, \ell_{11}^6 = \frac{\sqrt{v} \, Z(\Gamma) \, \ell_{11}^6 }{R} \sim \ell_{11}^6 \left(\frac{R}{\ell_{11}}\right)^{d/2}
\end{equation}
where $d$ is the highest degree of the charge $\Gamma$ (i.e. the dimension of
the associated brane: 0, 2, 4, or 6).  Thus in the limit $\ell_{11} \rightarrow
0$ we see that $m \rightarrow 0$ and the kinetic terms for the vector multiplets
vanish.  The scaling of the FI term also depends on the charge of the center $\Gamma$
\begin{equation}\label{eqn:thetap2}
	\frac{\ell_{11}^3 \, \theta_p}{\ell_s} = \sqrt{R \ell_{11}^3}\, \textrm{Im}(e^{-i\alpha} Z(\Gamma_p)) \sim p^0 R^2  + {\mathcal O}(\ell_{11})
\end{equation}
with the constant piece proportional to the D6 charge $p^0$.  So exactly as in
\cite{Boer2008b} the FI term (which maps to the constants in the integrability
equations (\ref{eqn:int})) survives only if a given center carries D6-charge.

Note that the limit we have taken is actually an M-theory limit as we are forced
to take the M-theory radius to infinity in Plank units $R/\ell_{11} \rightarrow
\infty$.  Thus the associated near-horizon region is AdS$_3\times$S$^2$.  If we
instead wished to stay in IIA we would have to take $\ell_{11}\rightarrow 0$
keeping $R/\ell_{11}$ fixed but this would send $R\rightarrow 0$ (i.e. $\ell_s
\rightarrow 0$ but keeping $g_s$ finite).  As evident from (\ref{eqn:thetap2})
this would also send the FI term to zero as $R^2$.

\bibliographystyle{JHEP}
\bibliography{refs}

\end{document}